\documentclass{emulateapj}
\newcommand{\HI}{\ion{H}{1}}
\newcommand{\HII}{\ion{H}{2}}
\slugcomment{Accepted to ApJ on August 3 2010}
\shorttitle{MILKY WAY DISK-HALO \HI\ CLOUDS}
\shortauthors{Ford et al.}
\begin{document}

\title{Milky Way Disk-Halo Transition in \HI : Properties of the Cloud Population}
\author{H. Alyson Ford\altaffilmark{1,2,3},
  Felix J. Lockman\altaffilmark{4},
  N. M. McClure-Griffiths\altaffilmark{2}}
\altaffiltext{1}{Centre for Astrophysics and Supercomputing, Swinburne
University of Technology, Hawthorn, Victoria 3122, Australia;}
\altaffiltext{2}{Australia Telescope National Facility, CSIRO Astronomy \& Space Science, Epping, NSW 1710, Australia;}
\altaffiltext{3}{Department of Astronomy, University of Michigan,
Ann Arbor, MI 48109; haford@umich.edu}
\altaffiltext{4}{National Radio Astronomy Observatory, Green Bank, WV 24944.}

\begin{abstract}

  Using 21~cm \HI\ observations from the Parkes Radio Telescope's
  Galactic All-Sky Survey, we measure 255 \HI\ clouds in the lower
  Galactic halo that are located near the tangent points at $16.9
  \degr \leq l \leq 35.3 \degr$ and $|b| \lesssim 20\degr$. The
  clouds have a median mass of 700~M$_{\odot}$ and a median distance
  from the Galactic plane of 660~pc. This first Galactic quadrant
  (QI) region is symmetric to a region of the fourth quadrant (QIV)
  studied previously using the same data set and measurement criteria.
  The properties of the individual clouds in the two quadrants are
  quite similar suggesting that they belong to the same population, 
  and both populations have a line of sight cloud-cloud velocity dispersion of $\sigma_{cc}
  \approx 16$~km~s$^{-1}$.  However, there are three times as many
  disk-halo clouds at the QI tangent points and their scale height, at
  $h=800$~pc, is twice as large as in QIV.  Thus the observed line
  of sight random cloud motions are not connected to the cloud scale
  height or its variation around the Galaxy. The surface density of
  clouds is nearly constant over the QI tangent point region but is
  peaked near $R\sim4$~kpc in QIV.  We ascribe all of these differences
  to the coincidental location of the QI region at the tip of the
  Milky Way's bar, where it merges with a major spiral arm. The QIV
  tangent point region, in contrast, covers only a segment of a minor
  spiral arm.  The disk-halo \HI\ cloud population is thus likely tied
  to and driven by large-scale star formation processes, possibly
  through the mechanism of supershells and feedback.

\end{abstract}


\section{Introduction}

The atomic hydrogen (\HI ) in the Milky Way has long been known to lie in a
thin layer with a FWHM of a few hundred pc in the inner Galaxy
\citep{1957Schmidt}.  The layer, however, consists of multiple
components, each with a different scale height, and the densest and
coolest gas is more confined to the Galactic plane than the warmer,
more diffuse gas \citep{1975Baker, 1984Lockman, 1987Savage,
1990Dickey, 2009Savage}.  The connection between scale height and
physical temperature seems natural but is misleading, for temperature
alone is not sufficient to support any of the \HI\ components to their
observed height --- additional support is needed, most likely from
turbulence.  In the vicinity of the Sun it is plausible that the observed  turbulence is
sufficient to support the \HI\ layer \citep{1991Lockman, 2009Koyama},
but in the inner Galaxy, at a Galactocentric radius $R\approx 4$~kpc,
one \HI\ component has an exponential scale height of 400 pc
\citep{1990Dickey}, and would require a turbulent velocity dispersion
$\sigma_z > 70$ km s$^{-1}$ to achieve its vertical extent
\citep{2009Kalberla}.  Evidence for the existence of a medium with
these properties is contradictory \citep{1998Kalberla,2003Howk}.  
The \HI\ component with the largest scale height may be involved in circulation of 
gas between the disk and halo, and probably contains the majority of the
kinetic energy of the neutral ISM \citep{1985Kulkarni,1991Lockman}.

The discovery that the transition region between the Galactic disk and
halo in the inner Galaxy contains many discrete \HI\ clouds that have
a spectrum of size and mass, and that follow normal Galactic rotation
to several kpc from the plane \citep{2002Lockman}, changed the picture considerably.
  A significant fraction of the \HI\ far from the
Galactic plane may be contained in these clouds, which are clearly part of a disk population unrelated to high-velocity clouds.  They are much denser than their surroundings, but do not have  the density needed for
gravitational stability.   Because the \HI\ clouds are seen in a continuous 
distribution from  the Galactic disk up to several kpc into the halo
\citep{2006Stil,2002Lockman}, we will refer to them as disk-halo clouds.  Clouds with similar properties have been detected in the outer
Galaxy \citep{2006Stanimirovic, 2010Dedes}.  Their origin, lifetime,
evolution, and connection with other interstellar components is
unknown.  They might result from a galactic fountain
\citep{1976Shapiro,1980Bregman,1990Houck,2008Spitoni}, \HI\ shells and
supershells \citep{2006McClure-Griffiths}, or interstellar turbulence
(e.g., \citealt{2005Audit}).  The properties of this population are
currently not well determined.  A thorough understanding of the
clouds, their physical nature, and their role in the Galaxy is
therefore important for understanding the circulation of material
between the Galactic disk and halo, a critical process in the
evolution of galaxies.

In an earlier paper (\citealt{2008Ford}; hereafter Paper~I) we
presented an analysis of the \HI\ disk-halo cloud population in the
fourth Galactic quadrant of longitude (hereafter QIV) based on new
observations made with the Parkes Radio Telescope\footnote{The Parkes
Radio Telescope is part of the Australia Telescope which is funded by
the Commonwealth of Australia for operation as a National Facility
managed by CSIRO.}. Those data spanned $324.7\degr \le l \le
343.1\degr$ and $|b|\lesssim 20\degr$, within which about 400 \HI\
clouds were detected.  Analysis of a subset of 81 clouds whose
kinematics placed them near the tangent points, and thus at a known
distance, indicated that the QIV clouds have a line of sight
cloud-cloud velocity dispersion $\sigma_{cc} = 18$~km~s$^{-1}$, far
too low to account for their distances from the plane if $\sigma_z
\approx \sigma_{cc}$.

The Galactic All-Sky Survey (GASS; \citealt{2009McClure-Griffiths}),
from which the observations of Paper~I were drawn, covers all
declinations $\delta \leq 1\degr$ and thus a significant portion of
the Galactic plane in the first quadrant of longitude (hereafter QI).
In this paper we analyze the \HI\ in the QI region mirror-symmetric in
longitude to the QIV region analyzed in Paper~I.  This allows us to
study the variation of cloud properties with location in the Galaxy
using a uniform data set and a uniform set of selection criteria.
We select a sample of clouds whose kinematics place them near tangent points, 
allowing their properties to be determined reasonably well.  The QI tangent point clouds can be compared directly with the equivalent QIV tangent point sample from Paper~I. The QI cloud sample turns out to be quite large,  revealing 
trends that were only hinted at in earlier data.

We begin with a description of the data (\S \ref{sec:p2quadIregion}),
then present the observed and derived properties of the disk-halo
clouds that lie near the QI tangent points (\S \ref{sec:p2observedprops}
 and \S \ref{sec:p2derivedprops}). A simulation of the cloud
population is used to determine distance errors and to better
characterize both the spatial distribution of the clouds and their
kinematics (\S \ref{sec:p2analysis}). The properties and distribution
of the clouds detected within the QI and QIV regions are compared in
\S \ref{sec:propcomp}, revealing marked differences in numbers and distributions.  In \S \ref{sec:galpop} we
consider possible origins for the differences 
 and examine potential selection effects, concluding that the differences relate to large-scale Galactic structure.  After a discussion of the fraction of halo \HI\ that might be contained in clouds (\S \ref{sec:fraction}) 
we consider the hypothesis that disk-halo clouds could be the product of stellar feedback 
and superbubbles (\S \ref{sec:superbubbles}).  A summary discussion is in \S
\ref{sec:summary}.

\section{Disk-Halo Clouds in the First Quadrant}

\subsection{The Observational Data}
\label{sec:p2quadIregion}

The data used in this paper are from the Galactic All-Sky Survey, an
\HI\ survey of the entire southern sky to declination $\delta \leq
1\degr$ made with the Parkes Radio Telescope
\citep{2009McClure-Griffiths}. GASS is fully sampled at an angular
resolution of $16\arcmin$, covers $-400 \leq V_{\mathrm{LSR}} \leq
+500$~km~s$^{-1}$ at a channel spacing of 0.8~km~s$^{-1}$, and has a
rms noise per channel of $\Delta T_b \approx 60$~mK. The data analyzed
here are from the first release of the survey which has not been
corrected for stray radiation.  This should not significantly affect
measured cloud properties, however, for clouds by definition are
isolated spatially and kinematically, and cannot be mimicked by stray
radiation, which tends to produce broad spectral features that vary
slowly with position \citep{1980Kalberla, 1986Lockman}.

\begin{figure}
  \includegraphics[scale=0.475]{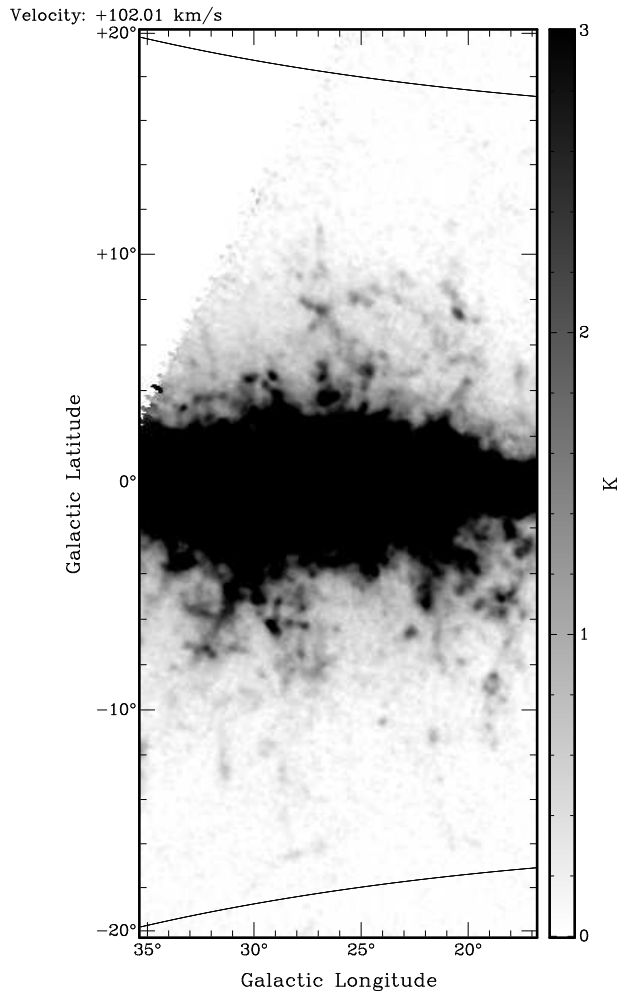}
  \caption[Channel map of GASS data  in quadrant I region.]
  {\HI\ intensity in a single GASS channel at
    $V_{\mathrm{LSR}}=102$~km~s$^{-1}$ over the region of the first
    Galactic quadrant studied here.  This area is symmetric in
    longitude to the fourth quadrant region studied in Paper~I
and covers the same latitudes.  The
    curved lines show the latitude boundaries of the region searched
    for clouds.   The
    lack of data in the top, left corner shows the declination
    limit of GASS.}
          \label{fig:lb}
\end{figure}

A region of the first quadrant of the Galaxy that spans $16.9\degr
\leq l \leq 35.3 \degr$ and $|b| \lesssim 20\degr$ was searched for
disk-halo \HI\ clouds.  This region is completely symmetric about
$l=0\degr$ to the region studied in Paper~I, i.e., it is at
complementary longitudes and thus covers an identical distance from
the Galactic center and distance from the Galactic plane. A channel
map of the GASS data for this region at
$V_{\mathrm{LSR}}=102$~km~s$^{-1}$ is presented in Figure
\ref{fig:lb}. Many clouds are apparent above and below the plane, and
many appear to be associated with loops and filaments, as were many
clouds in QIV. The lack of data in the top, left corner of Figure
\ref{fig:lb} shows the declination limit of GASS ($\delta \sim
1\degr$). Our comparisons with QIV are performed via simulations that
take this limit into account.

\subsection{The QI Tangent Point Sample}
\label{sec:tangentsample}

\begin{figure}
  \includegraphics[scale=0.6]{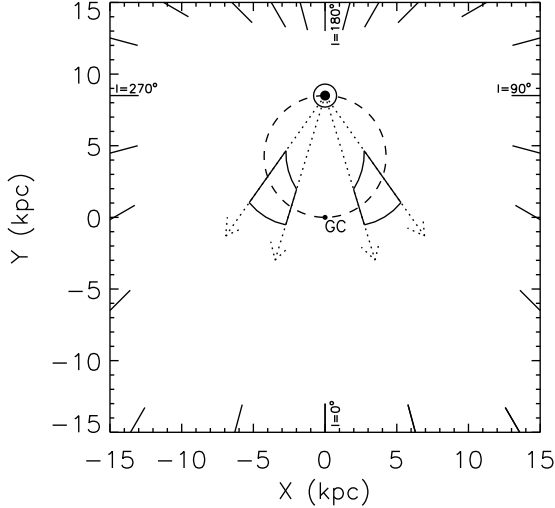}
  \caption{Schematic view of the QI and QIV tangent point regions of the Galaxy
    discussed in this paper.  The longitude boundaries of the regions
    are denoted by the dotted arrows and the locus of tangent points
    is the dashed circle that connects the Sun and the Galactic
    center. The solid lines enclose the area around the tangent points
    bound by the longitude limits and a kinematic distance $\pm
    18$~km~s$^{-1}$ about $V_{\mathrm{t}}$ for a flat rotation curve.
 Most of the clouds discussed here lie in these two areas.}
          \label{fig:tpdiagram}
\end{figure}

To define a sample of disk-halo clouds that has reasonably
well-determined properties, we limited the study of QI clouds to those
located near tangent points, where the largest velocity,  $V_{\mathrm{t}}$, occurs as the
line of sight reaches the minimum Galactocentric radius (see Figure
\ref{fig:tpdiagram}).  In the absence of random
motions, clouds in pure Galactic rotation cannot have
$V_{\mathrm{LSR}} > V_{\mathrm{t}}$, and all clouds with
$V_{\mathrm{LSR}}=V_{\mathrm{t}}$ would be located at tangent
points. However, random motions characterized by a cloud-cloud
velocity dispersion ($\sigma_{cc}$) can push a cloud's
$V_{\mathrm{LSR}}$ beyond $V_{\mathrm{t}}$. The deviation velocity,
$V_{\mathrm{dev}}\equiv V_{\mathrm{LSR}}-V_{\mathrm{t}}$, is a measure
of the discrepancy between a cloud's LSR velocity and the maximum
velocity expected from Galactic rotation in its direction. By focusing
on clouds with velocities beyond $V_{\mathrm{t}}$, i.e., with
$V_{\mathrm{dev}} \gtrsim 0$~km~s$^{-1}$ in QI, we isolate a
tangent point sample of clouds within a specific region of the Galaxy
whose volume is dependent on the cloud-cloud velocity dispersion
\citep{1979Celnik,2006Stil}.  Note that because we study clouds only
relatively close to the Galactic plane,
our definition of
$V_{\mathrm{dev}}$ assumes  corotation, but, unlike the definition used
for studies of high-velocity clouds \citep{1991bWakker}, does not include
presumptions about the thickness of the \HI\ layer.

We define the tangent point sample as those clouds with velocities
$V_{\mathrm{LSR}} \geq V_{\mathrm{t}} - 0.8$~km~s$^{-1}$, where
$0.8$~km~s$^{-1}$ accounts for one channel spacing. Clouds that meet
this velocity criterion must lie reasonably close to the tangent
point, and thus at a distance from the Sun $d_{\mathrm{t}} = R_0
\cos(l)/\cos(b)$ where $R_0 \equiv 8.5$ kpc.  The accuracy of the
adopted distances has been determined using the simulations of 
 \S \ref{sec:p2analysis}. Terminal velocities were taken from
the analysis of QI \HI\ by McClure-Griffiths \& Dickey (in
preparation) derived identically to their QIV work
\citep{2007McClure-Griffiths} used in Paper~I. At longitudes
outside the McClure-Griffiths \& Dickey range ($ l < 19\degr$), terminal
velocities were taken from the CO observations of \citet{1985Clemens}.

\begin{figure}
  \includegraphics[scale=0.475]{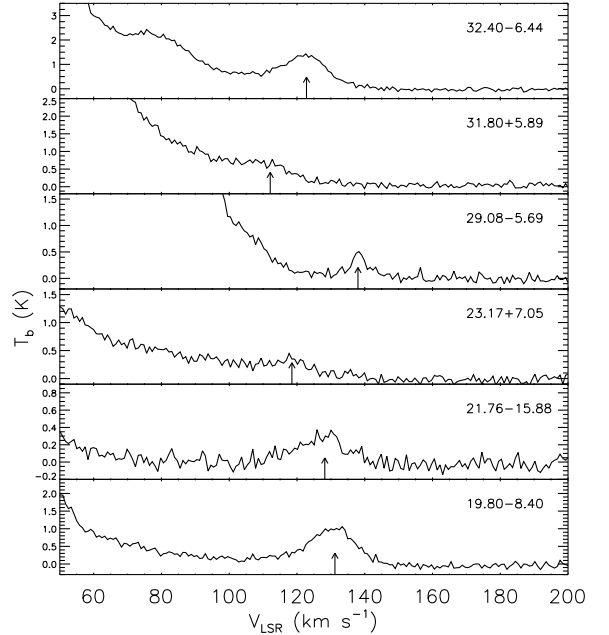}
  \caption[A random sample of GASS spectra from the quadrant I region.]
  {A random selection of spectra showing disk-halo clouds detected in the
    GASS QI data.}
          \label{fig:quadIspect}
\end{figure}

The procedure to detect and measure clouds in the QI tangent point
sample is identical to that for QIV and used the following two
criteria: (1) clouds must span 4 or more pixels and be clearly visible
over three or more channels in the spectra, and (2) clouds must be
distinguishable from unrelated background emission. Sample spectra of
clouds within QI are shown in Figure \ref{fig:quadIspect}.  Clouds are
blended and confused at low $V_{\mathrm{dev}}$ and near the Galactic
plane, and they become impossible to identify. We quantify this effect
and apply it to our simulations (\S \ref{sec:p2simulatedclouds}).
Paper~I gives further details on the selection criteria and search
method.

\subsection{Observed Properties of the QI Tangent Point Sample}
\label{sec:p2observedprops}

\begin{figure*}
\includegraphics[scale=0.34]{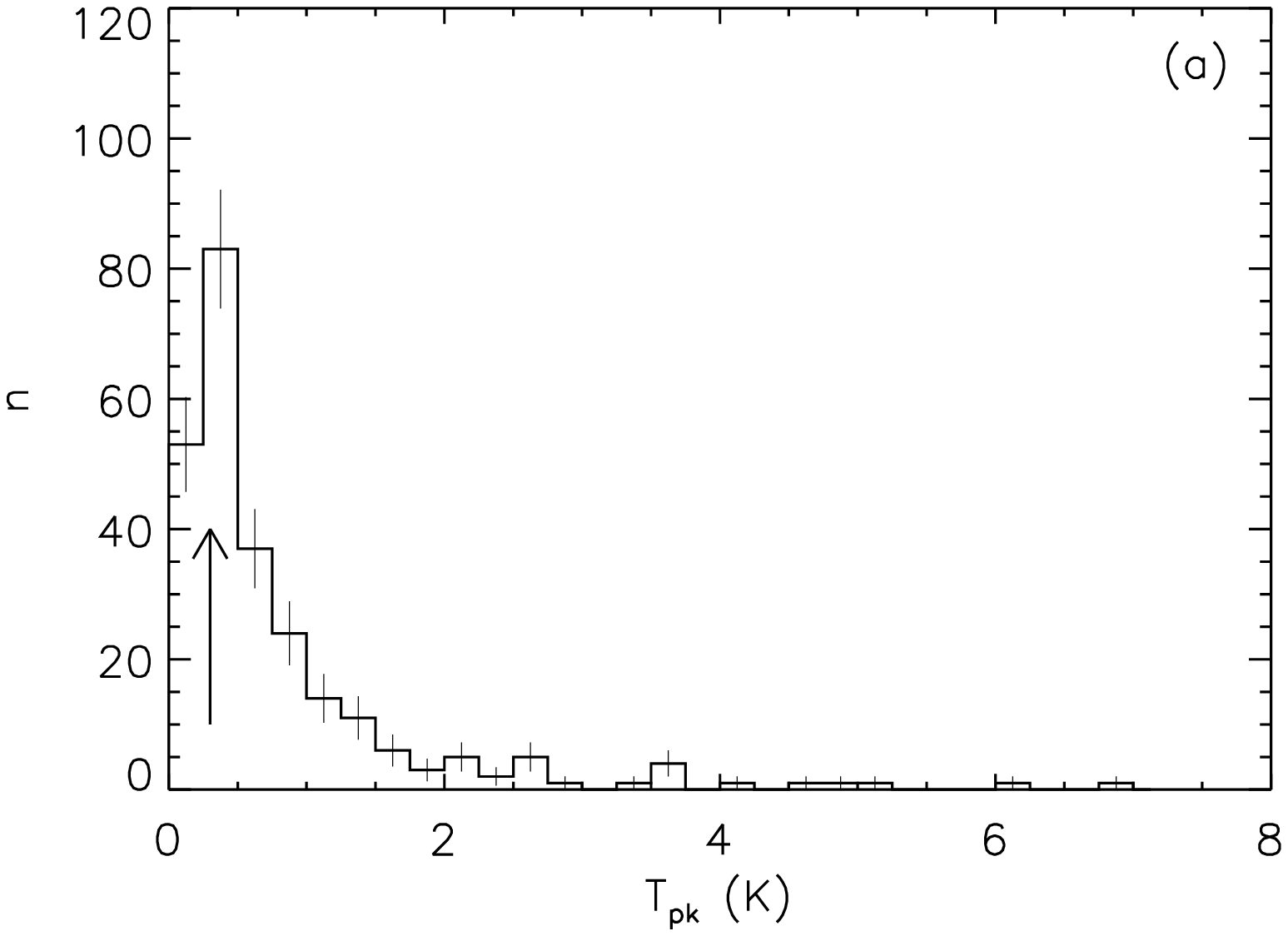}\includegraphics[scale=0.34]{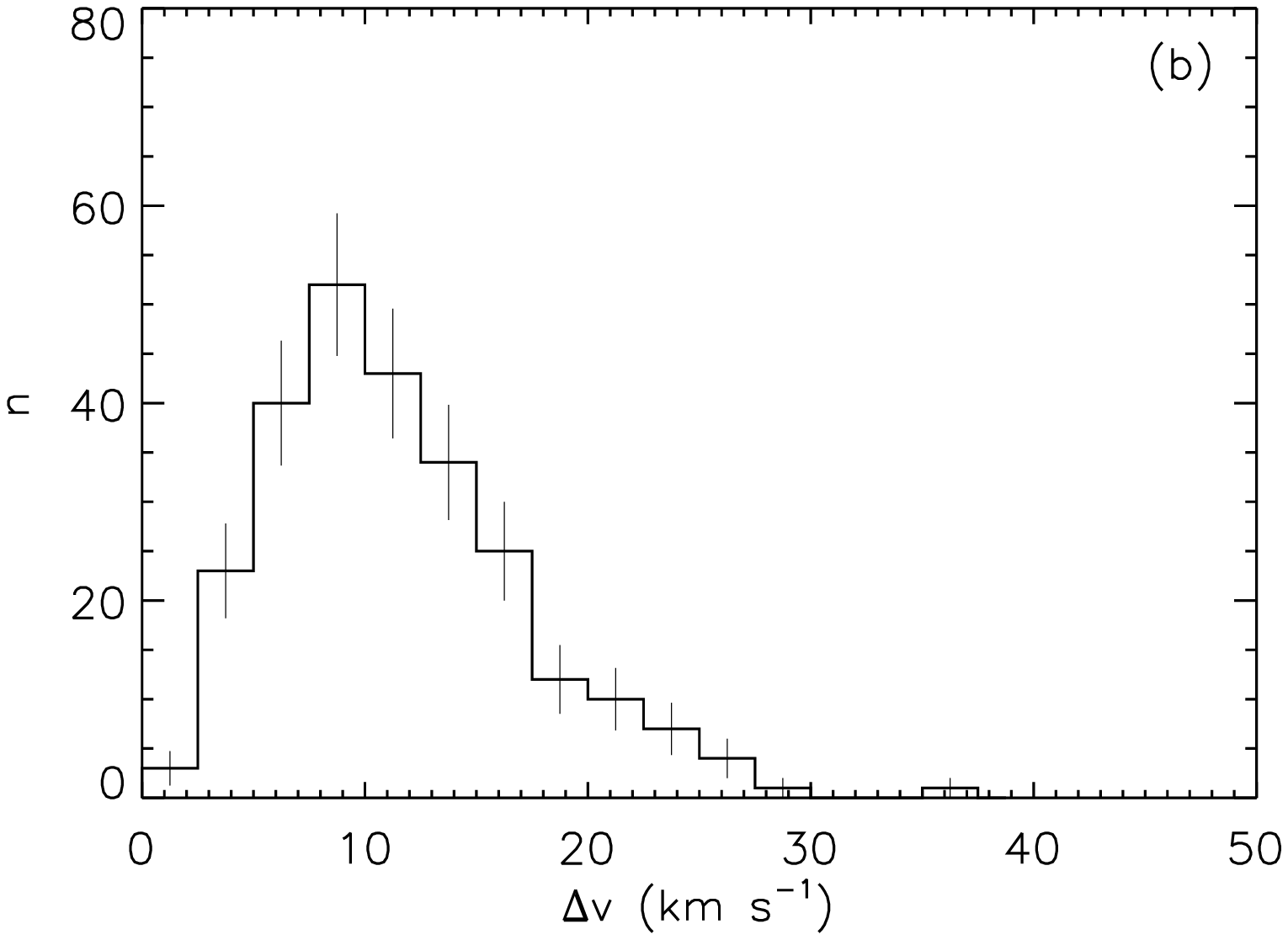}\includegraphics[scale=0.34]{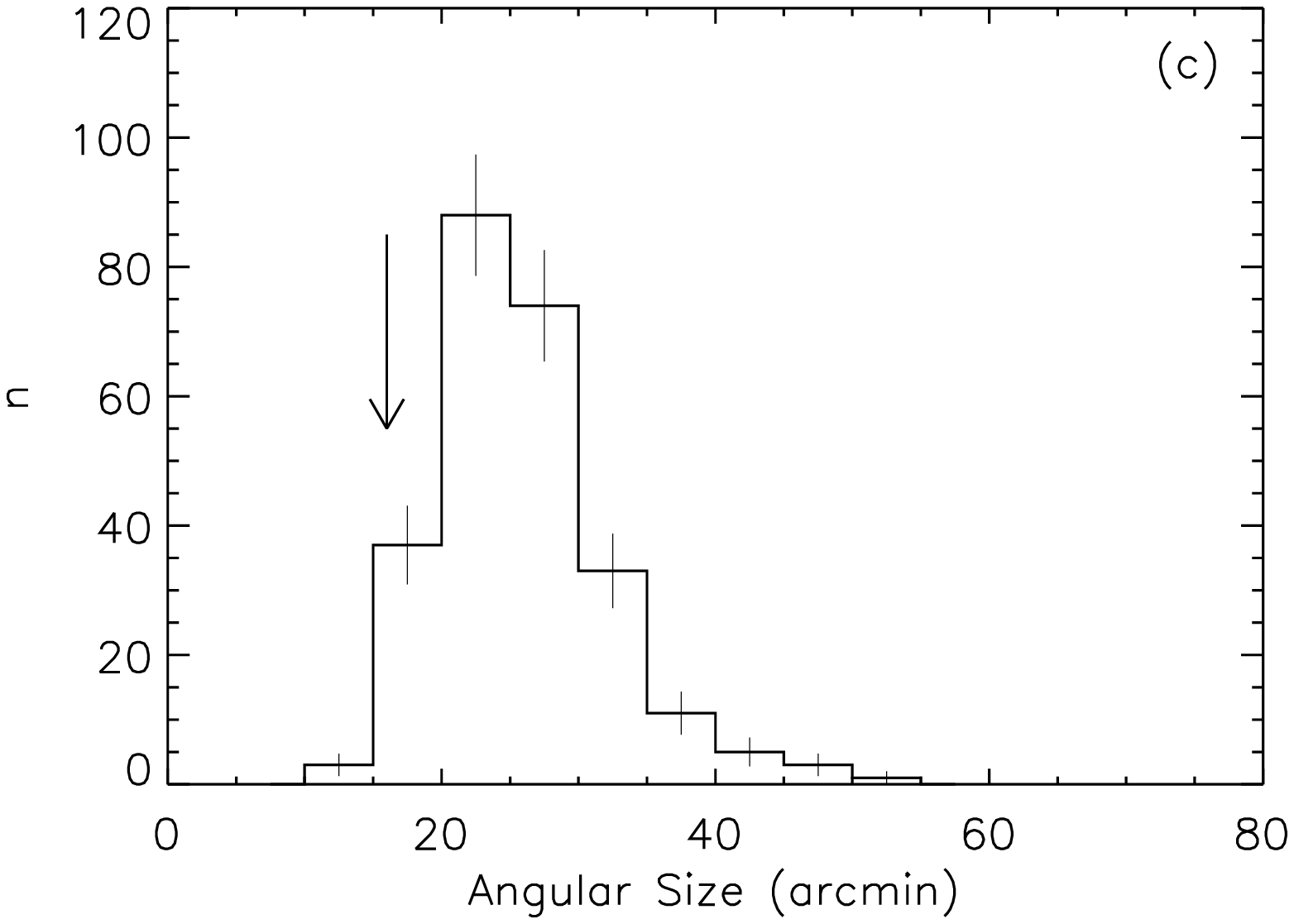}
\caption{\label{pTbhist} (a) Histogram of peak brightness temperature
  of the QI tangent point clouds.  The median $T_{\mathrm{pk}}=0.5$~K
  while the lower cutoff is due to the sensitivity limit. The arrow
  represents the $5\Delta T_b$ detection level. We assume  $\sqrt{N}$
  errors.  (b)\label{fwhmhist} Histogram of the FWHM of the velocity
  profile of the tangent point clouds. The median value is
  $10.6$~km~s$^{-1}$ and most profiles are well resolved.
  (c)\label{angularsizehist} Histogram of the angular size of the
  clouds, $(\theta_{\mathrm{maj}}\theta_{\mathrm{min}})^{1/2}$, where
  $\theta_{\mathrm{maj}}$ and $\theta_{\mathrm{min}}$ are from Table
  \ref{tab:Iquadobscatalogue}. The median angular diameter of the
  clouds is $25'$.  The spatial resolution limit of $16'$ is noted with
  the arrow.}
\end{figure*}

We detect and measure 255 \HI\ tangent point clouds in the QI
region. Their properties are presented in Table
\ref{tab:Iquadobscatalogue} and include Galactic longitude, $l$,
Galactic latitude, $b$, velocity with respect to the local standard of
rest, $V_{\mathrm{LSR}}$, peak brightness temperature,
$T_{\mathrm{pk}}$, FWHM of the velocity profile, $\Delta v$, peak \HI\
column density, $N_{HI}$, minor and major axes,
$\theta_{\mathrm{min}}$ and $\theta_{\mathrm{maj}}$, and the \HI\
mass, $M_{HI}d^{-2}$. These properties were determined analogously to
those in Paper~I. Histograms of peak brightness temperature, FWHM, and
angular size are presented in Figure \ref{pTbhist}. The clouds have
properties similar to those detected in QIV, with median values
$T_{\mathrm{pk}}=0.5$~K, $\Delta v=10.6$~km~s$^{-1}$, and angular size
$25'$. As many as 80\% of the clouds may be unresolved in at least one
dimension.
 
\begin{deluxetable*}{rrrrrrcc}
\tabletypesize{\footnotesize}
\tablewidth{0pt}
\tablecaption{Observed Properties of Tangent Point \HI\ Clouds in the Quadrant I Region\label{tab:Iquadobscatalogue}}
\tablehead{\colhead{$l$}&\colhead{$b$}&\colhead{$V_{\mathrm{LSR}}$}&\colhead{$T_{\mathrm{pk}}$\tablenotemark{a}}&\colhead{$\Delta v$}&\colhead{$N_{HI}$}&\colhead{$\theta_{\mathrm{min}}\times\theta_{\mathrm{maj}}$\tablenotemark{b}}&\colhead{$M_{HI}d^{-2}$\,\,\,\tablenotemark{c}}\\ 
\colhead{(deg)}&\colhead{(deg)}&\colhead{(km~s$^{-1}$)}&\colhead{(K)}&\colhead{(km~s$^{-1}$)}&\colhead{($\times 10^{19}$~cm$^{-2}$)}&\colhead{(arcmin $\times$ arcmin)}&\colhead{($M_{\odot}$ kpc$^{-2}$)}}
\startdata
$17.30$ & $-8.80$ & $135.9 \pm 2.6$ & $0.84$ & $12.5 \pm 1.3$ & $2.04 \pm 0.28$ & $16 \times 30$ & $9.8 $ \\ 
$17.44$ & $2.17$ & $139.8 \pm 4.7$ & $0.35$ & $14.0 \pm 3.0$ & $0.95 \pm 0.28$ & $18 \times 21$ & $6.2 $ \\ 
$17.60$ & $-11.21$ & $134.4 \pm 4.9$ & $0.41$ & $16.0 \pm 2.9$ & $1.27 \pm 0.32$ & $20 \times 22$ & $5.1 $ \\ 
$17.72$ & $3.94$ & $133.6 \pm 10.2$ & $0.21$ & $21.8 \pm 7.5$ & $0.89 \pm 0.44$ & $20 \times 34$ & $8.7 $ \\ 
$17.85$ & $-9.66$ & $131.7 \pm 1.7$ & $0.87$ & $8.2 \pm 1.0$ & $1.39 \pm 0.21$ & $20 \times 24$ & $7.7 $ \\ 
\enddata
\tablecomments{Table \ref{tab:Iquadobscatalogue} is published in its
  entirety in the electronic edition of the {\it Astrophysical
    Journal}. A portion is shown here for guidance regarding its form
  and content. Properties were determined analogously to those
  described in Paper~I.}  \tablenotetext{a}{Uncertainties in
  $T_{\mathrm{pk}}$ are $0.07$~K.}  \tablenotetext{b}{Uncertainties in
  the maximum angular extents are dominated by background levels
  surrounding the cloud and are assumed to be $25\%$ of the estimated
  values.}  \tablenotetext{c}{Mass uncertainties are dominated by the
  interactive process used in mass determination and are assumed to be
  $40\%$ of the estimated values.}
\end{deluxetable*}

\begin{figure}
  \includegraphics[scale=0.475]{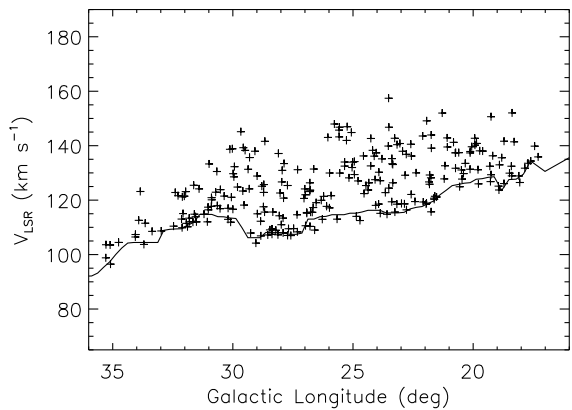}
\caption[Longitude vs. $V_{\mathrm{LSR}}$ of tangent point
clouds.]{Longitude--velocity diagram of clouds with $V_{\mathrm{LSR}}
  \gtrsim V_{\mathrm{t}}$ (crosses) within the QI region.  The
  terminal velocity curve (solid line) was derived from \HI\ and CO
  observations near the Galactic plane.  The steep decline in the
  number of clouds with $V_{\mathrm{LSR}} \gg V_{\mathrm{t}}$ (see
  also Figures \ref{fig:zvdev} and \ref{fig:GASS_Q14_vdev}) shows
  that the motions of these clouds are dominated by Galactic
  rotation. We searched all $V_{\mathrm{LSR}}\leq 300$~km~s$^{-1}$ but
  find no clouds with $V_{\mathrm{LSR}}> 160$~km~s$^{-1}$.}
\label{fig:lv}
\end{figure}

Figure \ref{fig:lv} shows the longitude vs. $V_{\mathrm{LSR}}$ of all
clouds from the tangent point sample within the QI region, along with
the adopted terminal velocity curve. Although we searched for clouds
at all velocities between $V_{\mathrm{t}}$ and
$V_{\mathrm{LSR}}=300$~km~s$^{-1}$, there were no clouds detected at
$V_{\mathrm{LSR}}>160$~km~s$^{-1}$, and no cloud has a velocity $>
50$~km~s$^{-1}$ beyond that allowed by Galactic rotation. Just as in
the QIV data of Paper~I, there is a steep decline in the number of
clouds beyond the terminal velocity indicating that the kinematics of
these clouds are dominated by Galactic rotation.

\subsection{Derived Properties of the QI Tangent Point Sample}
\label{sec:p2derivedprops}

In Table \ref{tab:Iquaddercataloguetangent} we present the cloud
properties that depend on the assumption that the clouds are located
at tangent points, which include the distance, $d$, Galactocentric
radius, $R$, distance from the Galactic plane, $z$, radius, $r$, and
physical mass of \HI, $M_{HI}$.  The deviation velocity,
$V_{\mathrm{dev}}\equiv V_{\mathrm{LSR}}-V_{\mathrm{t}}$, is also
presented, which shows the discrepancy between a cloud's LSR velocity
and the maximum velocity expected from Galactic rotation in its
direction. These properties were determined analogously to those for
the QIV clouds as described in Paper~I. Errors were also determined
analogously, with the uncertainty in the terminal velocity, $\delta
V_{\mathrm{t}}$, taken to be $3$~km~s$^{-1}$ for all clouds at $l >
19\degr$, where the terminal velocities were determined from \HI\
observations, and $9$~km~s$^{-1}$ for longitudes where the terminal
velocity was derived from CO observations. These adopted uncertainties
are identical to those used in Paper~I; the difference between \HI\
and CO uncertainties is expected to account for the granularity of
molecular clouds \citep{1978Burton,2007McClure-Griffiths}.  For
quantities whose uncertainties depend on distance, a simulated
population of clouds was used to determine these effects.

\begin{deluxetable*}{rrrrrrrrr}
\tabletypesize{\footnotesize}
\tablewidth{0pt}
\tablecaption{Derived Properties of Tangent Point \HI\ Clouds in the Quadrant I Region\label{tab:Iquaddercataloguetangent}}
\tablehead{
\colhead{$l$}&\colhead{$b$}&\colhead{$V_{\mathrm{LSR}}$}&\colhead{$V_{\mathrm{dev}}$}&\colhead{$d$}&\colhead{$R$\,\,\tablenotemark{a}}&\colhead{$z$}&\colhead{$r$}&\colhead{$M_{HI}$}\\
\colhead{(deg)}&\colhead{(deg)}&\colhead{(km~s$^{-1}$)}&\colhead{(km~s$^{-1}$)}&\colhead{(kpc)}&\colhead{(kpc)}&\colhead{(kpc)}&\colhead{(pc)}&\colhead{($M_{\odot}$)}}
\startdata
$17.30$ & $-8.80$ & $135.9\pm 2.6$ & $ 5.1 \pm 9.4$ & $ 8.2 \pm 0.6$ & $ 2.5^{+0.1}$ & $ -1.26 \pm 0.10$ & $ 26 \pm 5$ & $ 660 \pm 280$ \\
$17.44$ & $2.17$ & $139.8\pm 4.7$ & $ 6.6 \pm 10.2$ & $ 8.1 \pm 0.6$ & $ 2.5^{+0.1}$ & $ 0.31 \pm 0.02$ & $ 23 \pm 4$ & $ 410 \pm 180$ \\
$17.60$ & $-11.21$ & $134.4\pm 4.9$ & $ 2.6 \pm 10.3$ & $ 8.3 \pm 0.6$ & $ 2.6^{+0.1}$ & $ -1.61 \pm 0.12$ & $ 25 \pm 5$ & $ 350 \pm 150$ \\
$17.72$ & $3.94$ & $133.6\pm 10.2$ & $ 1.3 \pm 13.6$ & $ 8.1 \pm 0.6$ & $ 2.6^{+0.1}$ & $ 0.56 \pm 0.04$ & $ 30 \pm 6$ & $ 580 \pm 250$ \\
$17.85$ & $-9.66$ & $131.7\pm 1.7$ & $ 3.1 \pm 9.2$ & $ 8.2 \pm 0.6$ & $ 2.6^{+0.1}$ & $ -1.38 \pm 0.11$ & $ 26 \pm 5$ & $ 520 \pm 220$ \\
\enddata
\tablecomments{Table \ref{tab:Iquaddercataloguetangent} is published
  in its entirety in the electronic edition of the {\it Astrophysical
    Journal}. A portion is shown here for guidance regarding its form
  and content. Properties were determined analogously to those
  described in Paper~I.} 
 \tablenotetext{a}{Along a given line of
  sight, the smallest Galactocentric radius possible is at the tangent
  point. If the cloud is not located at the tangent point it must be
  farther away from the center and the error on $R$ must be positive.}
\end{deluxetable*}

\begin{figure}
  \includegraphics[scale=0.475]{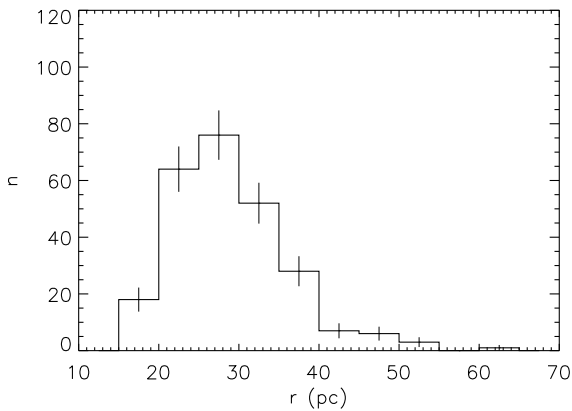}
  \caption[Histograms of physical radius.]
  {Histogram of physical radius of the QI tangent point clouds. The
    median radius is $28$~pc.  The spatial resolution limit of $16'$
    at the median cloud distance gives a limiting radius of
    $18$~pc. We assume $\sqrt{N}$ errors. }
\label{fig:p2radiushist}
\end{figure}

\begin{figure}
  \includegraphics[scale=0.475]{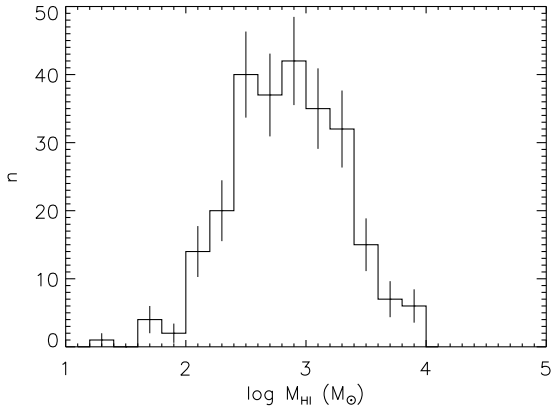}
  \caption[Histograms of physical mass.]
  {Histogram of the \HI\ mass of the tangent point clouds in the QI
    data set. The median mass is $\sim 700\ M_{\odot}$. We assume
    $\sqrt{N}$ errors.}
\label{fig:masshist}
\end{figure}

The distributions of radii and mass of the clouds are shown in Figures
\ref{fig:p2radiushist} and \ref{fig:masshist}. The median radius is
$28$~pc while the median mass in \HI\ is $700 M_{\odot}$. It is likely
that the individual values for the radii are overestimates, as a
significant fraction of clouds appear to be unresolved.  Confusion of
unrelated clouds may also increase the measured angular size of small
clouds. We used simulations to interpret the $V_{\mathrm{dev}}$, $z$,
and $R$ distributions and present the results in 
the following section.

\section{Analysis of  the QI Tangent Point Sample}
\label{sec:p2analysis}

\subsection{Simulation of the Cloud Distribution}
\label{sec:p2simulatedclouds}

As demonstrated in Paper~I, simulations can establish the fundamental
characteristics of an observed population, including the uncertainties
introduced by the assumption that all clouds with $V_{\mathrm{dev}}
\gtrsim 0$, i.e., $V_{\mathrm{LSR}} \gtrsim V_{\mathrm{t}}$, are at
the tangent point.  Indeed, without simulations it is difficult to accurately
derive quantities such as the surface density or scale height from the
tangent point cloud data.  Consider, for example, a population of clouds in a
small area at a tangent point in QI that have a characteristic cloud-cloud 
velocity dispersion $\sigma_{cc}$.  The ensemble has on average
$\langle V_{\mathrm{LSR}}\rangle = V_{\mathrm{t}}$, and exactly half of the clouds
will have $V_{\mathrm{LSR}} \geq V_{\mathrm{t}}$ and thus be included
in the tangent point sample.  An area along the line of sight somewhat
closer to us than the tangent point will have 
$\langle V_{\mathrm{LSR}}\rangle <
V_{\mathrm{t}}$, but owing to random motions, some fraction of the
clouds --- fewer than half --- will nonetheless have $V_{\mathrm{LSR}}
\geq V_{\mathrm{t}}$ and thus end up in the sample of
``tangent point'' clouds.  The number that do depends on the number of
clouds at the offset location and the cloud-cloud velocity dispersion.  
Because the change in $\langle V_{{\mathrm{LSR}}}\rangle$ with
distance from the Sun may be only 5-10 km s$^{-1}$ kpc$^{-1}$ over much of
the inner Galactic disk, a very large volume of the Galaxy must be
simulated to encompass all clouds likely to end up with forbidden
velocities.

 By determining the functions that best represent the observed data,
an understanding of the properties of the cloud population is obtained
that is otherwise unattainable. We simulated a population of clouds to
represent the observed tangent point population within QI, applying
the same $l$, $b$ and $V_{\mathrm{LSR}}$ selection criteria as for the
observed population. A cut was also applied to account for the
declination limit of the data.  The functional form for the cloud
distribution is identical to that adopted for QIV:
\begin{equation}
  n(R,z) = \Sigma(R)\exp\left({-\frac{|z|}{h}}\right),
\end{equation}
where $\Sigma(R)$ is the surface density in kpc$^{-2}$, $h$ is the
exponential scale height, and $R$ and $z$ are the cylindrical
coordinates. $\Sigma(R)$ is composed of $16$ independent bins of width
$0.25$~kpc, spanning $R=2.5$ to $6.5$~kpc.

The velocities of the simulated clouds were derived assuming a flat
rotation curve where $\Theta=\Theta_0=220$~km~s$^{-1}$ with a
random line of sight component drawn from a Gaussian with a dispersion
$\sigma_{cc}$. Corotation is assumed: there is no variation in
Galactic rotational velocity with distance from the plane. This issue
is discussed in \S \ref{sec:corotation}. Over the first and second
quadrants, $5\times 10^4$ clouds were generated of which 1954 lay
within the defined $l$, $b$, $V_{\mathrm{LSR}}$, and $\delta$ range of
the QI clouds. This set was then normalized to compare directly with
the observed population and Kolmogorov-Smirnov (K-S) tests were
performed to estimate the quality of the fit between observation and
simulation.  Results of the fits to the distributions are presented in
\S\S \ref{sec:p2veldispersion}--\ref{sec:p2verthist}. We defer a
discussion of the implications of these findings along with a detailed
comparison between the first and fourth quadrant clouds until \S
\ref{sec:propcomp}.

To determine the uncertainties introduced by the assumption that all
clouds with $V_{\mathrm{dev}} \gtrsim 0$ are located at tangent points
at a distance $d_{\mathrm{t}} = R_0 \cos(l)/\cos(b)$, we calculated the
fractional distance error of the simulated clouds as a function of
deviation velocity, longitude and latitude. As shown by the
simulations presented in Paper~I, clouds with increasingly forbidden
velocities have smaller distance uncertainties, i.e., are more likely
to be actually located at $d_{\mathrm{t}}$.  
This can be understood as follows.  The cloud in QI with the highest
forbidden velocity has $V_{\mathrm{dev}}\approx 42$~km~s$^{-1} \approx 3\sigma_{cc}$ (see below).  
From a group of $\sim 100$ clouds at the tangent points with a random velocity
of $\sigma_{cc}$ we would expect to find $\sim 1$ such cloud. However, at a
location away from the tangent point, where the $\langle V_{\mathrm{LSR}}\rangle$ is much lower,
a cloud would have to have a more extreme random velocity,
$\gg 3\sigma_{cc}$ to have the identical $V_{\mathrm{dev}}$.  
Thus the greater the $V_{\mathrm{dev}}$, the more likely, on average, that a cloud has 
come from a population with high $V_{\mathrm{LSR}}$ and is thus near the tangent point.
There is also a slight
dependence of the distance uncertainties on longitude, but no
dependence on latitude. Likewise, there is a longitude and deviation
velocity dependence for the uncertainty in Galactocentric distance. 

\begin{figure}
  \includegraphics[scale=0.475]{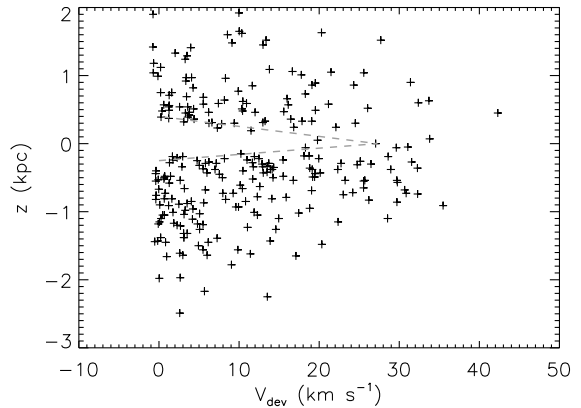}
  \caption{Distance from the Galactic plane as a function of deviation
    velocity of QI tangent point clouds.  Confusion limits the detectability of clouds 
    that have low values of $V_{\mathrm{dev}}$ and are near the plane. Dashed lines 
    represent the cutoffs that were applied to the simulations to reflect the effects of 
    this confusion. Note the decrease in number of clouds as a function of 
    $V_{\mathrm{dev}}$ at all $z$, showing that the disk-halo clouds follow Galactic rotation.}
\label{fig:zvdev}
\end{figure}

Confusion effects are a strong function of proximity of clouds to the
bulk of Galactic \HI : clouds will increasingly be blended and missed
at lower heights and lower values of $V_{\mathrm{dev}}$. This is
obvious from Figure \ref{fig:lb}. The confusion manifests itself in
the $z$ vs. $V_{\mathrm{dev}}$ plot of Figure \ref{fig:zvdev} as the
triangular region near the Galactic plane that is devoid of detected
clouds. To account for these effects, we have parameterized the
boundaries of the confusion-affected region by the dashed lines shown
in Figure \ref{fig:zvdev}, and apply these cutoffs to both the
observed and simulated data to compare the two as directly as
possible. This is in contrast to how we addressed these effects with
the QIV data, where we simply omitted all clouds with $|b|\leq
2\degr$, as the number of tangent point clouds in QIV was too small to
justify a more detailed cutoff.

\subsection{Cloud-Cloud Velocity Dispersion}
\label{sec:p2veldispersion}

The simulated population of QI clouds that best represents the observed first quadrant 
population has a random cloud-cloud velocity component drawn from a Gaussian with a 
dispersion $\sigma_{cc}=14.5$~km~s$^{-1}$  (with a K-S test probability of $51\%$ that the observed
and simulated distributions were drawn from the same distribution).  Values of $\sigma_{cc}$ 
between $14$ and 
$15.5$~km~s$^{-1}$ also provide acceptable fits to the measured $V_{\mathrm{dev}}$, having 
K-S test probabilities $\ge 15\%$.

\subsection{Radial Surface Density}
\label{sec:p2raddist}

The amplitude of each radial bin, $\Sigma(R)$, was optimized to best fit the observed 
longitude distribution of the tangent point clouds by minimizing the Kolmogorov-Smirnov $D$ 
statistic (the maximum deviation between the cumulative distributions) using Powell's
algorithm \citep{1992Press}. The fits were optimized for three different initial estimates 
of $\Sigma(R)$, and as they all converged on a similar solution, we adopted the mean of the 
three solutions as the best fit to the data.

\begin{figure}
  \includegraphics[scale=0.475]{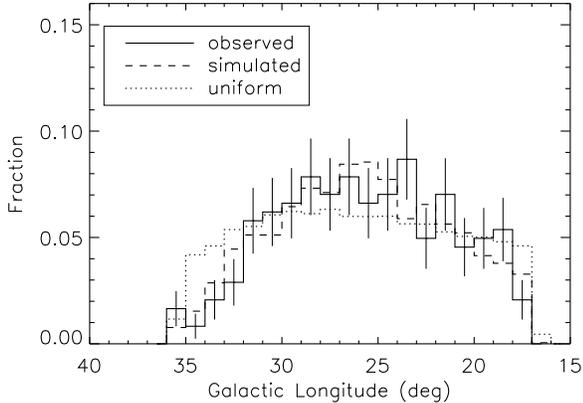}
  \caption[Longitude distribution of observed, simulated and a uniformly distributed population of clouds.]
  {Longitude distribution of observed (solid line) and simulated
    (dashed line) QI clouds, with a uniform surface density population
    overlaid (dotted line). The uniformly distributed population of
    clouds closely resembles the observed population except at the highest longitudes,  suggesting that the tangent point clouds
    within the QI region are distributed rather evenly. We
   assume $\sqrt{N}$ errors.}
         \label{fig:lhist}
\end{figure}

The longitude distribution of both the observed and simulated population of clouds, along 
with that derived from a population of clouds with a uniform surface density, is shown in 
Figure \ref{fig:lhist}. The K-S test probability is 88\% that the observed and simulated 
distributions were drawn from the same distribution. Such an unusually high K-S test probability 
is not surprising in this case, as the parameters of the simulation were fine-tuned to 
reproduce the observed longitude distribution. The probability that the observed and a uniformly
distributed population were drawn from the same distribution is $3\%$, so the observations 
are marginally inconsistent with being drawn from a population of clouds with a uniform surface 
density. In this case most of the discrepancy arises from the bins with the 
highest longitude.  

\begin{figure}
  \includegraphics[scale=0.475]{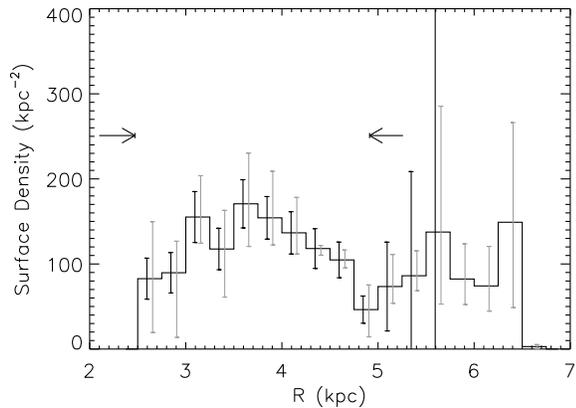}
  \caption[Radial surface density of tangent point clouds in quadrant I.]
  {Radial surface density distribution of the simulated population of
  QI  clouds.  The black error bars represent the Poisson error in the
    number of clouds expected to fall into the observed sample, which
    determined the ability of the observed data to constrain the
    simulated distribution.  The grey error bars represent the
    range in values for the 3 different starting points used in the
    minimization.  Error bars are large in the right-most bins because
    very few clouds at these radii have high enough random velocities
    such that their $V_{\mathrm{LSR}}>V_{\mathrm{t}}$.  
    No simulated clouds within the largest radial bins meet the selection criteria of the tangent 
    point sample. The bins without black error bars are therefore unconstrained by the observed 
    cloud population. Arrows represent the tangent points at the longitude
    boundaries of the QI region.}
          \label{fig:surfacedens}
\end{figure}

The most likely radial surface density distribution is shown in Figure \ref{fig:surfacedens}.  
Although the area we studied included tangent points only over $2.5 \leq R \leq 4.9$ kpc, the 
simulations indicate that a few clouds at larger $R$ are expected to have random velocities
that, when added to their rotational velocity, give them a $V_{\mathrm{LSR}} > V_{\mathrm{t}}$ 
and thus place them in the tangent point sample.

\subsection{Vertical Distribution}
\label{sec:p2verthist}

\begin{figure}
  \includegraphics[scale=0.475]{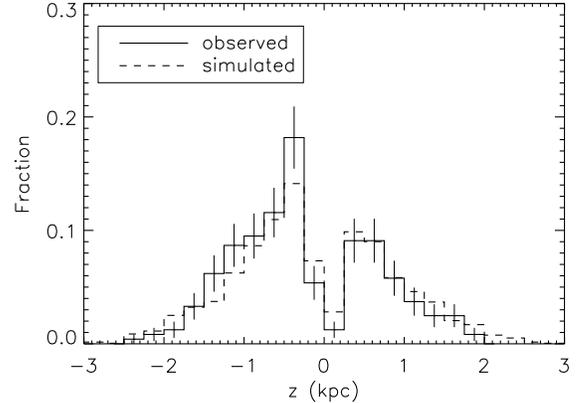}
  \caption[Vertical distribution of observed and simulated clouds.]
  {Vertical distribution of the observed QI (solid line)
    and simulated (dashed line) population of clouds.
 The observed
    population is well represented by a simulated population with an
    exponential scale height of $800$~pc.  Note that inclusion of a
    simulated ``confusion-cutoff'' (Figure \ref{fig:zvdev}) does a
    reasonable job of reproducing the decrease of clouds near the
    plane. We assume $\sqrt{N}$ errors.}
          \label{fig:zhist}
\end{figure}

The vertical distribution of the tangent point clouds is best
represented by an exponential with a scale height $h=800$~pc (see
Figure \ref{fig:zhist}). Because K-S tests are most sensitive to the
differences near the median of the distribution, we tested both $n(b)$
and $n(|b|)$: the former gives higher weight to clouds closer to the
plane, while the latter weighs more highly the higher latitude clouds.
Also, fitting to $n(b)$ is sensitive to symmetries about the plane
while fitting to $n(|b|)$ is not. For $h=800$~pc, the K-S test probability that the
observed and simulated clouds were drawn from the same vertical
function is $61\%$ when comparing against $|b|$, but not acceptable
when comparing against $b$. Scale heights between $700$ and $850$~pc
were acceptable for the distribution of $|b|$, while scale heights
between 950 and $1100$~pc were consistent with the distribution of $b$
values, with 1000 pc being the best fit, having a probability of
$24\%$.  This discrepancy likely indicates that our low-latitude
confusion cutoff (Figure \ref{fig:zvdev}) is not conservative enough.
As this effects the distribution of $b$ more strongly than of $|b|$,
we adopt the preferred value from the latter comparison, $h = 800$~pc.

\section{Comparison of Cloud Populations in the Two Quadrants}
\label{sec:propcomp}

\subsection{Trends in Physical Properties}
\label{sec:proptrends}

The physical properties of disk-halo tangent point \HI\ clouds from QI and QIV are summarized in Table
\ref{tab:p2comptable}.  Individual clouds in both quadrants of
the Galaxy have similar  properties, which suggests that they belong to
the same population and probably have similar origins and evolutionary
histories.  We defer an analysis of the physical properties of the
clouds to another paper; here we note just a few trends that give
important insight into the nature of the disk-halo clouds.

\begin{deluxetable}{lcccc}
  \tablecaption{Properties of Tangent Point Disk-Halo Clouds in GASS \label{tab:p2comptable}}
  \tablehead{ & \multicolumn{2}{c}{Median} & \multicolumn{2}{c}{90\% Range}\\
    \colhead{Parameter} & \colhead{QI} & \colhead{QIV} & \colhead{QI} & \colhead{QIV}}
  \startdata
  $T_{\mathrm{pk}}$ (K) & $0.5$ & $0.5$& $0.2 \rightarrow 2.7$  & $0.2 \rightarrow 2.1$ \\
  $\Delta v$ (km~s$^{-1}$) & $10.6$ & $10.6$ & $4.2 \rightarrow 23.2$  & $4.4 \rightarrow 22.3$\\
  $N_{HI}$ ($\times 10^{19}$ cm$^{-2}$) & $1.0$ & $1.0$ & $0.2 \rightarrow 6.2$  & $0.3 \rightarrow 6.5$\\
  Angular Size ($'$) & $25$ & $28$ & $17 \rightarrow 38$ & $21 \rightarrow 44$\\
  $r$ (pc) & $28$ & $32$ &  $< 19 \rightarrow 44$ & $< 23 \rightarrow 50$\\
  $M_{HI}$ ($M_{\odot}$) & $700$  & $630$  & $130 \rightarrow 4100$ & $120 \rightarrow 4850 $\\
  $d$ (kpc) & $7.7$ & $7.7$ & $7.2 \rightarrow 8.1$ & $7.2 \rightarrow 8.1$ \\
  $R$ (kpc) & $3.7$ & $3.6$ &  $2.7 \rightarrow 4.6$ & $2.9 \rightarrow 4.6$ \\
  $|z|$ (pc) & $660$ & $560$  & $190 \rightarrow 1640$ & $320 \rightarrow 1690 $ \\
  \enddata
  \tablecomments{Median values of the tangent cloud properties, where
    the number of tangent point clouds is 255
    for QI and 81 for QIV. Most properties have a large scatter about the median in
    all samples, as demonstrated by the $90\%$ range.  The angular size is
$(\theta_{\mathrm{maj}}\theta_{\mathrm{min}})^{1/2}$. }
\end{deluxetable}

First, as the gas mass required for a cloud to be gravitationally bound is 
$M\approx r\Delta v^2/G$, where $r$ is the radius, $\Delta v$ is the FWHM,
and $G$ is the gravitational constant, the clouds
fail to be self-gravitating by several orders of magnitude.  This was
noted in the discovery of the disk-halo cloud population, and applies not only
to individual clouds, but also to dense clumps within clouds revealed
in high-resolution observations \citep{2002Lockman,2009Pidopryhora}.

\begin{figure}
  \includegraphics[scale=0.475]{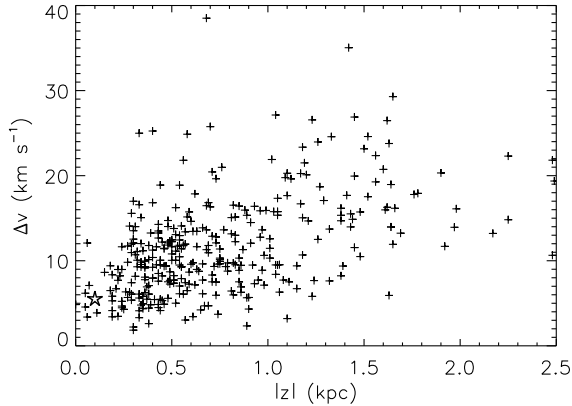}
  \caption[FWHM as a function of height.]
  {FWHM of the \HI\ profile, $\Delta v$, as a function of distance
    from the Galactic plane for all tangent point clouds from QI and
    QIV (crosses). The star at $|z|=0.1$~kpc and $\Delta
    v=5.5$~km~s$^{-1}$ shows the median values from the
    \citet{2006Stil} sample of clouds near the Galactic plane.
    Linewidths tend to increase with distance from the Galactic
    plane.}
  \label{fig:fwhmz}
\end{figure}

Second, there is strong evidence that clouds farther from the Galactic
plane have larger linewidths than clouds nearer the plane (Figure
\ref{fig:fwhmz}). This confirms previous suggestions of the trend
\citep{2002Lockman, 2006Stil, 2008Ford}. Moreover, given its
continuity with $|z|$, it does not appear to arise from the
superposition of separate populations of clouds, but supports the
hypothesis that it reflects pressure variations throughout the halo
and the fundamental role that pressure has in controlling the phases
of the neutral ISM \citep{1995Wolfire,2009Koyama}. Indeed, many clouds at
$|z|\sim 1$ kpc exhibit a two-phase structure with broad and narrow
line components \citep{2005Lockman}.

The clouds must be either pressure-confined or transitory features, as
they are not gravitationally bound. Without higher resolution data
than those presented here, we are unable to measure the pressure of
the clouds, so cannot comment directly on the first possibility. If
they are transitory features, however, their lifetimes must be
significantly less than both the thermal evaporation timescale, which
is $\ga 100$~Myr \citep{2006Stanimirovic} and the internal dynamical
time ($\sim 3$~Myr for the median cloud radius and linewidth). While
the first of these timescales is consistent with them evolving on a
free-fall timescale ($\sim 30$~Myr from 1 kpc, based on the vertical
potential of \citealt{1997Benjamin}), as might be expected if they are
lifted or ejected into the halo from the disk or if they form above
the disk and rain down, they expand and disperse much more quickly 
unless there is some form of confinement.

\subsection{Cloud Location and Numbers}
\label{sec:numbersanddist}

A summary of the properties of disk-halo \HI\ clouds at the QI and QIV tangent points is presented in Table \ref{tab:p3comptable}.  Although the first and fourth quadrant regions span the same Galactocentric
radii and vertical distances from the plane, differing only by being
located on opposite sides of the Sun-center line, there is a striking
difference in the number of detected tangent point clouds: 255 in QI compared to only 81 in QIV.

\begin{deluxetable*}{lccccc}
  \tablecaption{Summary of Disk-Halo \HI\ Cloud Distributions in QI and QIV\label{tab:p3comptable}}
  \tablehead{ & \multicolumn{2}{c}{QI} & \multicolumn{2}{c}{QIV} &\\
    \colhead{Parameter} & \colhead{Value} & \colhead{Best Fit} & \colhead{Value} & \colhead{Best Fit} & \colhead{Consistent?}}
  \startdata
  $n_{\mathrm{tp}}$ & $255$ & & $81$ & & No\\
  $\sigma_{cc}$ (km~s$^{-1}$) & & $14-15.5$ & & $16-22$ & Yes\\
  $h$ (pc) & & $700-850 $ & & $300-500$ & No\\
  $\Sigma(R)$ & & $\sim$uniform  & & $\sim$concentrated & No
  \enddata
  \tablecomments{The quantity $n_{\mathrm{tp}}$ is the number of clouds detected at the 
    tangent points. Properties derived from simulations are also presented. The number 
    of clouds is strikingly different between quadrants, as are the
    vertical scale heights and radial surface density distributions.
    The cloud-cloud velocity dispersions, however, are similar in both quadrants.}
\end{deluxetable*}

\begin{figure}
  \includegraphics[scale=0.475]{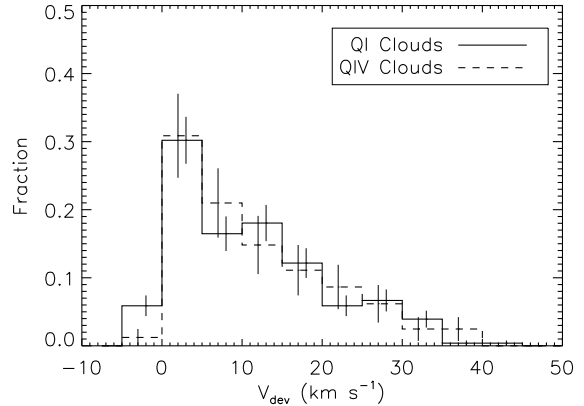}
  \caption[Vdev for the entire GASS sample]
          {Distribution of observed $V_{\mathrm{dev}}$ for the QI tangent point clouds (solid 
            line) and the symmetric $-V_{\mathrm{dev}}$ for the QIV clouds (dashed line). The 
            K-S probability is $76\%$ of them having been drawn from the same distribution, and 
            suggests that the cloud-cloud velocity dispersion in both quadrants is identical.
            This is supported by the simulations, which show that both populations have 
            distributions consistent with  $\sigma_{cc}=16$~km~s$^{-1}$.
            We assume $\sqrt{N}$ errors.}
          \label{fig:GASS_Q14_vdev}
\end{figure}

Values of $V_{\mathrm{dev}}$ for clouds in QI and QIV are similar (Figure \ref{fig:GASS_Q14_vdev}), 
with a K-S probability of $76\%$ of
having been drawn from the same distribution, suggesting that the
cloud-cloud velocity dispersion in both quadrants is identical. This
conclusion is supported by the simulations, which are consistent with
$\sigma_{cc}\approx 16$~km~s$^{-1}$ in both quadrants. The cloud
distributions, however, differ significantly in both the vertical and
radial directions.

\subsection{Vertical Distribution of Disk-Halo \HI\ Clouds}
\label{sec:p3scaleheight}

\begin{figure}
  \includegraphics[scale=0.475]{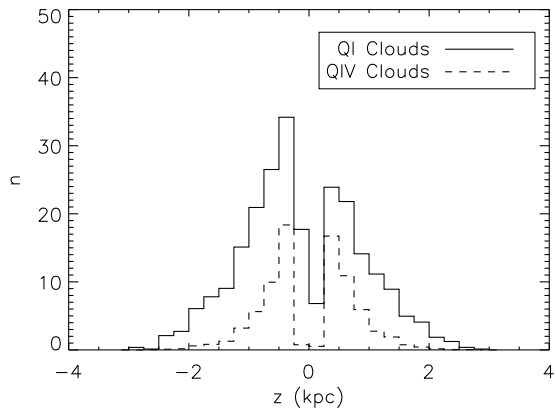}
  \caption[Derived vertical distributions of \HI\ clouds in quadrants I and IV.]
  {Comparison of the derived vertical distributions of disk-halo  \HI\ clouds in QI (solid line) and QIV (dashed line)
after taking into account selection criteria and confusion.
  Both distributions are well represented by an
    exponential, but with a scale height of $h=800$~pc in QI and only
    $h=400$~pc in QIV.}
            \label{fig:nofz}
\end{figure}

The vertical distributions of the tangent point clouds derived from
the simulations are shown in Figure \ref{fig:nofz}. These
distributions properly account for selection effects that artificially
skew the appearance of the observed distributions, particularly near the plane.  The scale heights differ between
quadrants, with $h=400$~pc in QIV compared with $h=800$~pc in QI, as
do the acceptable ranges of the scale height: in QIV 
acceptable fits range from $h=300$ to $500$~pc while in QI 
 acceptable fits are between $h=700$ and $850$~pc. We tested
to see if this result could occur because of differences in confusion
or asymmetries about the plane, but find it quite robust: there is no
overlap in acceptable ranges of $h$ between the QI and QIV tangent point cloud samples.

The origin of the scale height is unclear. If the clouds have a
vertical random motion equal to their line of sight random motion,
i.e.  $\sigma_{z} = \sigma_{cc}$, their scale height at $R \approx 4$
kpc would be less than 100 pc.  To achieve their observed distances
from the plane  at
$R=3.8$~kpc the clouds would have to have $\sigma_{z} > 60$
km~s$^{-1}$ in QIV and $> 95$ km~s$^{-1}$ in QI in the model potential of \citet{2007Kalberla}. Thus, given that $\sigma_{cc}$ is essentially identical
in both quadrants, {\it not only is it impossible for the derived
  vertical scale height of clouds to arise from motions with a
  magnitude of the cloud-cloud velocity dispersion $\sigma_{cc}$, but
  $\sigma_{cc}$ cannot be connected to the scale height.}  We will
discuss likely explanations for the scale height in a later section.

\subsection{Radial Surface Density}
\label{sec:p3radsurfdens}

\begin{figure}
  \includegraphics[scale=0.475]{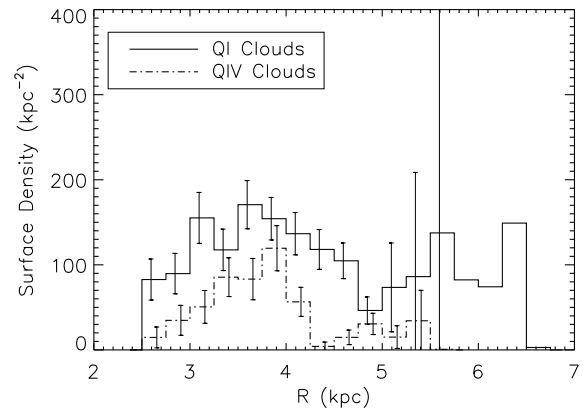}
 \caption[Derived radial surface density
 distributions of disk-halo \HI\ clouds.]
 {Comparison of the derived radial surface density distribution of
   disk-halo \HI\ clouds within the QIV region (dashed-dot line) and
   the QI region (solid line). There are $\approx 3$ times as many
   disk-halo clouds at the tangent points in QI as in QIV, and the
   radial distributions differ significantly as well.  Error bars
   denote the Poisson error in the number of clouds expected to fall
   into each bin. The right-most bins have no error bars because no clouds at 
   these $R$ are expected to meet the criteria of the tangent point sample. These
   bins are therefore unconstrained by the observed cloud population.}
          \label{fig:radsurfdens}
\end{figure}

The radial surface density of clouds derived from the simulations,
$\Sigma(R)$, is given in Figure \ref{fig:radsurfdens}.
It shows not only the much larger 
 number of clouds in QI than in QIV, but that the
shape of the distributions are fundamentally
different. The surface density in QI is relatively uniform, while the
surface density in QIV is concentrated, and declines rapidly at $R >
4.2$~kpc.  The difference is statistically robust. 
 We have compared the shapes of $\Sigma(R)$ within the
well-constrained $2.5 \leq R \leq 5$~kpc bins of the two quadrants,
rescaling them to have the same number of clouds so as to compare only
the shape, by calculating the $\chi^2$ of their difference compared to
the null hypothesis that the shapes are identical. The null hypothesis is strongly ruled
out with $\chi^2=31.5$ over 10 degrees of freedom, or a probability of
$0.05\%$. 

\section{Origin of the QI-QIV Asymmetries}
\label{sec:galpop}

Even though the individual disk-halo \HI\ clouds have virtually identical properties in the two quadrants, there are three major asymmetries in the populations: 1) There are three times as many clouds in the QI volume as in the identical volume of QIV. 2)  The scale height of the QI clouds is twice that of QIV clouds. 3) The radial surface density distributions are not at all alike.
In this section we consider possible explanations for these differences.

\subsection{Selection Effects}

\subsubsection{Instrumental Effects, Cloud Selection, Confusion}

\begin{figure}
  \includegraphics[scale=0.475]{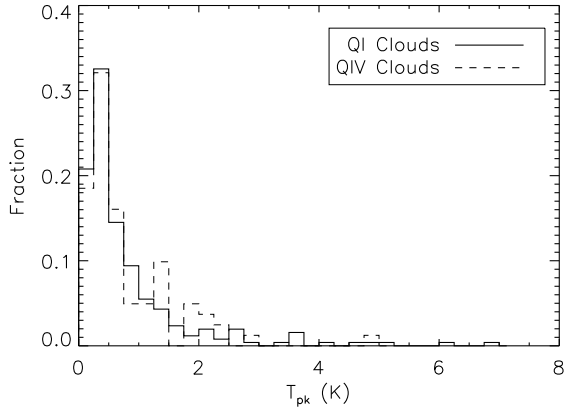}
  \caption{Comparison of the normalized distribution of observed peak line temperatures in 
    the QI (solid line) and QIV (dashed line) tangent point cloud samples.
    The sensitivity limits in QI and QIV are identical.}
  \label{fig:tpkhistcompobs}
\end{figure}

\begin{figure}
  \includegraphics[scale=0.475]{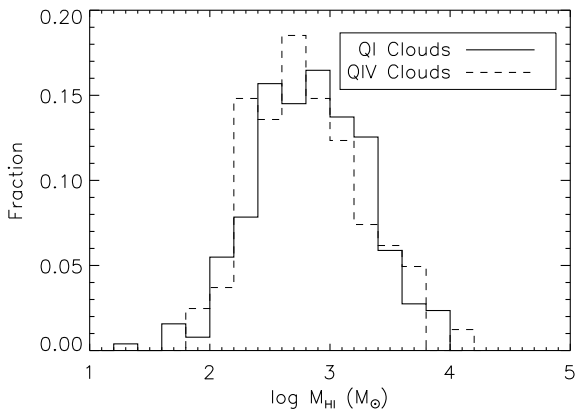}
  \caption{Comparison of the normalized distribution of cloud masses in the QI (solid line) and
    QIV (dashed line) tangent point cloud samples. There is no bias toward lower mass clouds 
    in QI.}
  \label{fig:mhihistcompobs}
\end{figure}

The GASS \HI\ data set has uniform instrumental properties.  Although the QIV clouds 
were measured from an early version of the data it differs little from the final version, 
and in no significant way that would produce a large  discrepancy between QI and QIV.  One 
of us (H.~A.~F.) developed the cloud identification and measurement techniques, and produced 
the cloud catalogues. Every attempt was made to apply the selection criteria uniformly to 
the two regions. The strongest evidence for a lack of bias in the cloud selection process 
is given by the near identity of individual cloud properties in the two regions.  This is 
apparent from Table \ref{tab:p2comptable}, and also from Figures \ref{fig:tpkhistcompobs} and 
\ref{fig:mhihistcompobs}, which show that the distribution of peak line intensity and cloud 
mass is essentially identical for the tangent point clouds in QI and QIV. If there were 
differences in the detection level or noise in the two data sets we would expect to see, 
for example, more clouds with small $T_{\mathrm{pk}}$ or small \HI\ masses in QI.  This 
does not occur.  We conclude that changes in survey sensitivity, or cloud identification 
and selection criteria, are not the source of the QI-QIV differences.

One difference between the QI and QIV data is that the area surveyed in QI is $\sim15\%$ 
smaller due to the declination limit of GASS (see Figure \ref{fig:lb}). However, this 
will only 
cause us to {\it underestimate} the difference in the number of clouds. We note that our 
simulations take this effect into account and the radial surface densities, vertical 
distributions, and cloud-cloud velocity dispersions are therefore all unaffected. 

\begin{figure}
  \includegraphics[scale=0.475]{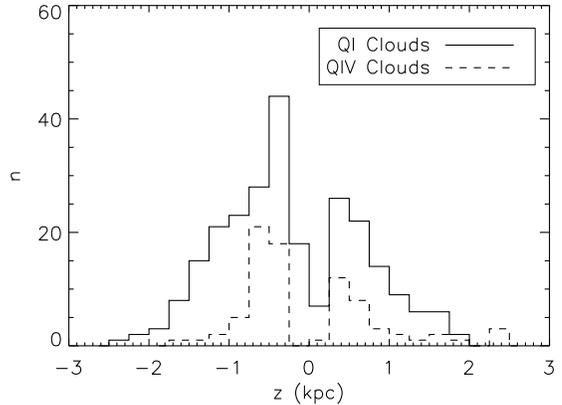}
  \caption{Comparison of the observed distance from the plane of \HI\
    clouds in the QI (solid line) and QIV (dashed line) tangent point
    samples. There are more clouds in QI than QIV at virtually every
    distance from the plane, eliminating any variation in confusion as a
    source of the QI-QIV differences.}
            \label{fig:zhistcompobs}
\end{figure}

Near the Galactic plane, or at low values of $|V_{\mathrm{dev}}|$,
 clouds blend with each other and with unrelated emission making it
 impossible to identify and measure them.  It is conceivable that the
 cloud population in QI is less confused than in QIV, allowing more
 clouds to be detected.  If, for example, the QI clouds had a
 larger $\sigma_{cc}$ than the QIV clouds then more of them would lie at
 V$_{LSR} > V_t$ where they might be less confused with unrelated \HI.
 This might also give an apparent increase in scale height.  Table
 \ref{tab:p3comptable} shows, however, that $\sigma_{cc}$ is nearly
 identical in the two regions and if anything, is somewhat smaller in
 QI than in QIV.  Moreover, Figure \ref{fig:zhistcompobs} shows that
 there are more clouds observed in QI at nearly every distance from
 the Galactic plane, whereas confusion is important only at $|z|
 \lesssim 0.5$ kpc.  Confusion cannot be the source of the QI-QIV
 differences.

\subsubsection{Kinematic Selection Effects}

\begin{figure}
  \includegraphics[scale=0.475]{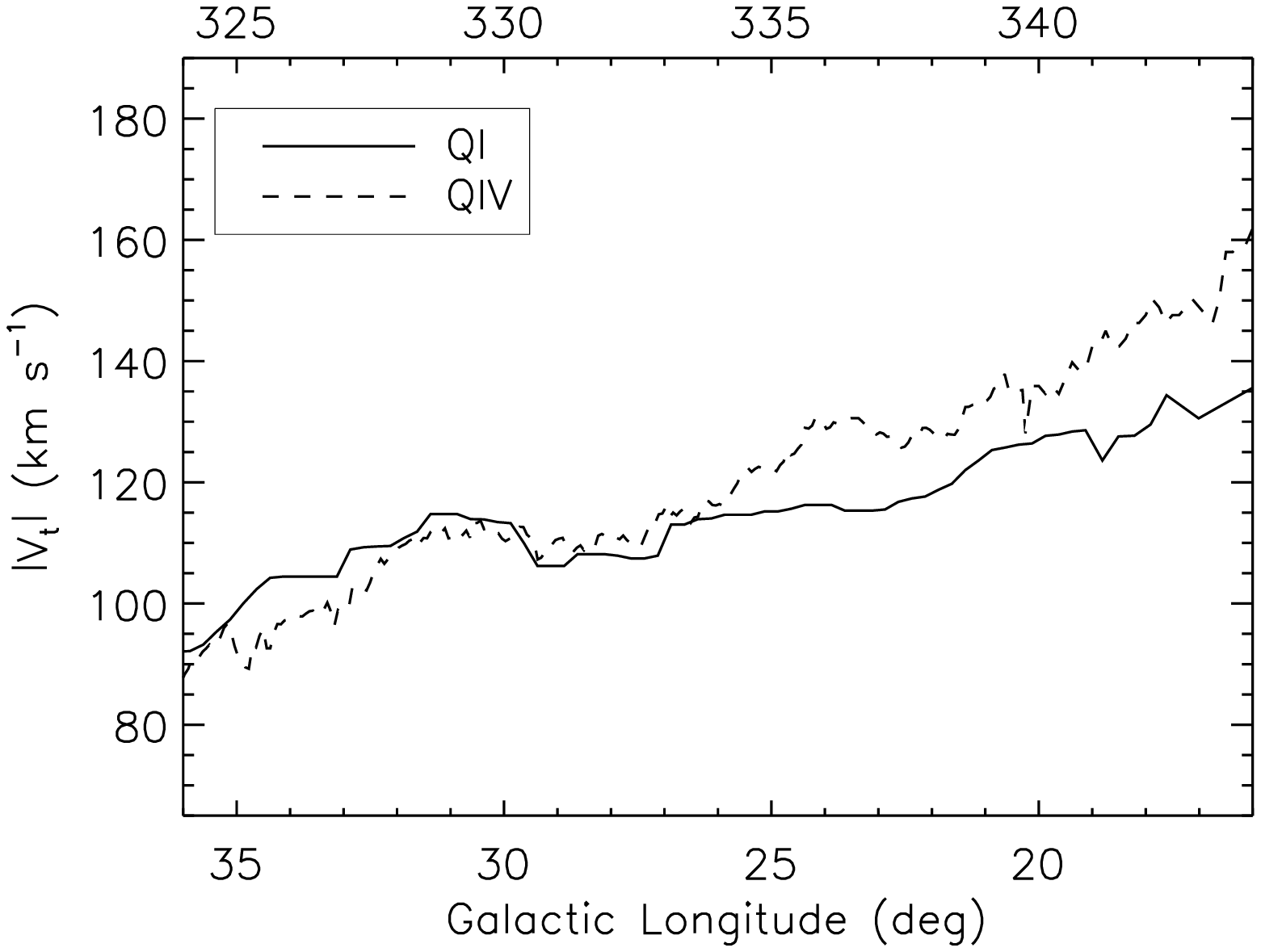}
  \caption{Adopted terminal velocity curves for QI and QIV. In QI, the
    terminal velocities were taken from the analysis of \HI\ by
    McClure-Griffiths \& Dickey (in preparation). At longitudes
    outside the McClure-Griffiths \& Dickey range ($l<19\degr$),
    terminal velocities were taken from the CO observations of
    \citet{1985Clemens}. In QIV, the terminal velocities were taken
    from the analysis of \HI\ by \citet{2007McClure-Griffiths}, while
    for $l \geq 339\fdg7$ we used terminal velocities determined by
    \citet{2006Luna}.}
  \label{fig:vtcurvecomp}
\end{figure}

The QI tangent point clouds were chosen from those with
$V_{\mathrm{LSR}} \gtrsim V_{\mathrm{t}}$, where $V_{\mathrm{t}}$ is
derived at each longitude from measurements of \HI\ or CO made close
to the Galactic plane.  Different authors use different methods to
derive $V_{\mathrm{t}}$, and we were careful to use values that treat
both quadrants uniformly (see \S \ref{sec:tangentsample}).  Our adopted
values of $V_{\mathrm{t}}$ are shown in Figure \ref{fig:vtcurvecomp}.
The QI values lie below those of QIV at about half of the
longitudes.  The existence of  large-scale deviations from symmetry in
the terminal velocities (and possibly in the rotation curve itself) was 
an early discovery of Galactic \HI\ studies \citep{1962Kerr}, and the reason that we
chose to use measured values of $V_{\mathrm{t}}$ to derive the cloud
samples rather than a theoretical rotation curve.  Nonetheless, it is
true that if for some reason the kinematics of the disk-halo cloud
population is actually symmetric about the Galactic center, our choice
of an asymmetric $V_{\mathrm{t}}$ curve would artificially inflate the
number of clouds in QI compared to QIV.

Could this be the cause of the differences we detect?  To explore this
we have calculated the number of QIV tangent point clouds that would
be found if the QI rather than the QIV form of $V_{\mathrm{t}}$ were
used.  This change would increase the number of QIV tangent point
clouds from 81 to 130, still a factor of two fewer than in QI.  In
another test, we selected clouds in QI using for $V_{\mathrm{t}}$ at
each longitude the maximum value from Figure \ref{fig:vtcurvecomp}.
This lowered the number of clouds in the QI sample from 255 to 180,
but still left it at twice the number as in QIV.  

To make the number of tangent point clouds equal in QI and QIV would
require the values of $V_{\mathrm{t}}$ to be systematically in error
by $>20$ km~s$^{-1}$ over at least $20\degr$ of longitude in one
quadrant only.  \citet{2008Levine} have independently derived terminal
velocities from \HI\ data over part of the longitude range of
interest.  Their values have, on average, a lower magnitude than those
we use, reflecting a difference in the adopted definition of
$V_{\mathrm{t}}$.  Nonetheless, their analysis shows an asymmetry
similar to that in Figure \ref{fig:vtcurvecomp}, and an average
difference between QI and QIV similar to the values we use.

We conclude that the difference in the disk-halo tangent point cloud sample between QI
and QIV is highly unlikely to arise from kinematic selection effects.
It is important to note as well that simply equalizing the numbers of
clouds in the two quadrants does not remove the substantial difference
in scale height (Figure \ref{fig:nofz}) or radial distribution
(Figure \ref{fig:radsurfdens}). We can find nothing in our observations
or analysis that would erroneously create differences of the observed
magnitude.  

\subsubsection{The Assumption of Corotation}
\label{sec:corotation}

Throughout this work we have assumed that there is no dependence of
the Galactic rotation curve on distance from the Galactic plane. Other
galaxies, however, show evidence for a systematic lag in rotation of
$10$--$20$~km~s$^{-1}$~kpc$^{-1}$ in both ionized and neutral gas that
begins perhaps 1 kpc from the plane and is of unknown origin
\citep{2005Fraternali, 2005Rand, 2010Marinacci}. Evidence for
departures from corotation in the Milky Way is scant
\citep{2007Pidopryhora}.  We defer a full analysis to a subsequent
paper, but note that because most of the clouds discussed here lie at
$|z|<1$ kpc, a lag may be difficult to detect. If
a significant number of the disk-halo clouds do lag somewhat behind
Galactic rotation, then our analysis has underestimated their numbers
at the tangent point, with the result that the true scale height, and
the true surface density, will be larger than given here, especially
in QI where the population extends much farther from the plane than in
QIV.

\subsection{Differences Caused by Galactic Structure}
\label{GLIMPSE}

\begin{figure}
  \includegraphics[scale=0.385]{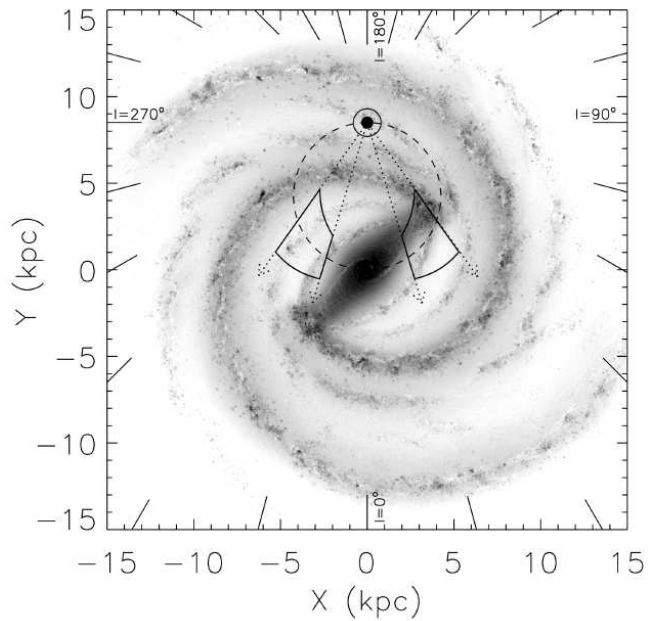}
  \caption[Milky Way stellar spiral structure with Quadrant I and IV regions overlaid.]
  {Artist's conception of the Milky Way as derived from GLIMPSE survey
    infrared data. The longitude boundaries of the QI and QIV regions
    are denoted by the dotted arrows while the locus of tangent points
    is represented by the dashed circle. The solid lines enclose the
    area around the tangent points bound by our longitude limits and a
    kinematic distance $\pm 18$~km~s$^{-1}$ away from
    $V_{\mathrm{t}}$ from Figure \ref{fig:tpdiagram}.  The two regions studied here, though selected
    to be symmetric with respect to the Sun-center line, actually
    sample two very different parts of the Galaxy. The QI area covers
    the near tip of the Galactic bar where a major spiral arm originates, while
    the QIV area contains only a minor arm. The artist's conception
    image is from NASA/JPL-Caltech/R. Hurt (SSC-Caltech).}
          \label{fig:GLIMPSE_image}
\end{figure}

Given that the difference in the numbers, scale height and surface
density are real, might they be carrying information on some
large-scale feature of the Milky Way?  Figure
\ref{fig:GLIMPSE_image} suggests that this is the case.  It shows an
idealization of the Galaxy with the outline
of the regions of the first and fourth quadrants from which the
tangent point clouds are selected (Figure \ref{fig:tpdiagram}). 
This depiction of the Galaxy
incorporates recent findings from the Galactic Legacy Infrared
Mid-Plane Survey Extraordinaire (GLIMPSE; \citealt{2003Benjamin}), a
survey that was conducted using the Spitzer Space Telescope. The
GLIMPSE data suggest that the spiral structure of the Galaxy, which
had originally been thought to consist of four major arms, is
dominated by two arms (the Scutum-Centaurus and Perseus arms) that
extend from each end of a central bar. There are also two minor arms
(the Norma and Sagittarius arms), located between the major arms. The
artist's conception incorporates the GLIMPSE findings on the location
of the bar and spiral arms \citep{2005Benjamin}, as well as recent
VLBI determinations of parallactic distance to some \HII\ regions
(e.g., \citealt{2009Reid}), the discovery of the far side of the 3-kpc
arm \citep{2008Dame}, and other evidence relevant to the overall
pattern of star formation in the Galaxy (R.~A. Benjamin, private
communication). The solid lines enclose areas within the longitude
limits of our study, with a line of sight extent equivalent to
$|V_{\mathrm{LSR}}| = |V_{\mathrm{t}}| - 18$~km~s$^{-1}$, i.e. the
$\sim 1\sigma$ volume around the tangent point for a flat rotation
curve.

In the GLIMPSE view of the Milky Way, the QIV tangent point region covers a rather
sparse section of the Galaxy through which only a segment of the
minor Norma spiral arm passes, while the QI region lies on a much
richer portion of the Galaxy where the near end of the Galactic bar
merges with the beginning of the major Scutum-Centaurus arm.  
The radial surface density distributions of the disk-halo
clouds mirror the spiral structure in both regions, suggesting a relation between 
the clouds and spiral features.
The association of the disk-halo clouds with regions of star
formation, and specifically with the asymmetry caused by the Galactic
bar, offers the only solution we can find to the asymmetries in the
disk-halo cloud distributions.

There is observational evidence that there has been a significant
burst of star formation near where the Scutum-Centaurus arm meets the
end of the bar, in the form of observations of multiple red supergiant
clusters in the area ($l\sim 29\degr$; e.g., see
\citealt{2006Figer,2007Davies,2009Alexander,2009Clark}). \citet{2003Motte}
even suggest that the \HII\ region complex W43, located at $l\sim
31\degr$, is a ``ministarburst.''  Other galaxies also have increased
star formation occurring where spiral arms meet the bar ends
\citep{1993Phillips}. We find a plethora of disk-halo \HI\ clouds
in the region where the Scutum-Centaurus arm extends from the bar end
in the first quadrant, three times more than in the fourth quadrant
region, suggesting that the number of clouds and their scale heights 
are proportional to the
amount of star formation. 

We can, however, find no detailed correlation between disk-halo clouds and other 
tracers of star formation.   Neither molecular clouds \citep{1988Bronfman}, nor 
HII regions \citep{2004Paladini}, show a strong QI-QIV asymmetry like the
disk-halo clouds, or have a radial surface density distribution with a
similar shape. 
Methanol masers detected in the 6.7 GHz emission line are known tracers of 
high-mass star formation \citep{2008Xu}, and some are formed in very early
protostellar cores \citep{2005Minier}. Methanol masers that have been detected 
by \citealt{2005Pestalozzi,2007Pandian,2007Ellingsen,2008Xu} and
\citealt{2009Cyganowski} within $\pm 30$~km~s$^{-1}$ of
$V_{\mathrm{t}}$ show a factor
2.2 excess  in the QI region compared with QIV, not as large
as the factor of 3.2 difference in disk-halo \HI\ clouds, but larger
than in any other population of which we are aware.  However, the
maser numbers peak at longitudes $\sim 30\degr$ on either side of
the Galactic center, a distribution not at all like the tangent point
\HI\ clouds.

We conclude that the numbers of disk-halo clouds correlate with the amount of star formation on the 
largest scales, but not with tracers of star formation in detail.

\section{Fraction of Galactic Extraplanar \HI\ in Disk-Halo Clouds}
\label{sec:fraction}

The new data allow estimation of the fraction of the total
extraplanar \HI\ that is in the form of the disk-halo clouds, through 
comparison  with previous derivations of $n_{HI}(z)$ in
the inner Galaxy (we cannot use the current release of GASS data to estimate $n_{HI}(z)$ 
as these data have not been corrected for stray radiation).  Over the first longitude quadrant,
\citet{1990Dickey} fit the vertical distribution of \HI\ along the
tangent points with a three-component model, one of which is an
exponential with a scale height of 400 pc.  This component contains
$1.6 \times 10^{20}$ cm$^{-2}$ through a full disk.  There are about
100 QI disk-halo clouds per kpc$^{2}$ (Figure \ref{fig:surfacedens}),
which at a median \HI\ mass of 700 $M_{\odot}$ sums to $10^{19}$~cm$^{-2}$ 
through a full disk, only about 5\% of the total $N_{HI}$
in the \citet{1990Dickey} exponential component. The cloud counts,
however, are very deficient near the plane because of confusion, and
the total number of disk-halo clouds could be 2-3 times larger than we
are able to identify.  Even so, it appears that the discrete disk-halo
\HI\ clouds identified in the GASS data account for no more
than perhaps 10-20\% of all \HI\ far from the plane in QI. 

There are many \HI\ structures above the disk which did not fit our
strict criteria for inclusion in the catalog of ``clouds'' but which
still contain significant amounts of \HI.  Also, there are certainly 
many small clouds detected at higher resolution \citep{2002Lockman} 
but missed in the GASS study.  The existence and extent
of a truly diffuse \HI\ halo component in the inner Galaxy thus remains
uncertain.  This topic will be addressed in a subsequent work when the
GASS survey data corrected for stray radiation becomes available.

\section{Disk-Halo Clouds as the Product of Stellar Feedback and Superbubbles}
\label{sec:superbubbles}

In Paper~I we suggested that the disk-halo clouds could be a result of
superbubbles and material that has been pushed up from the disk, as
the presence of many loops and filaments were clearly visible within
the QIV data, and many clouds appeared to be associated with these
structures.  Superbubbles 
can lift significant amounts of disk gas several kpc into the halo 
\citep{2009deGouveiaDalPino}.  They 
are quite common in areas of star formation
\citep{2009Kalberla}, so the presence of such structures would also be
expected in the QI region. The Ophiuchus superbubble is an excellent
example of such a structure: it is an old superbubble ($\sim 30$~Myr)
that lies within the first quadrant region and is above a section of
the Galaxy containing many \HII\ regions, including W43
\citep{2007Pidopryhora}. This superbubble is at a distance of $\sim
7$~kpc and is capped by a plume of \HI\ at $\sim 3.4$~kpc above the
plane. Many \HI\ features have been observed to be affiliated with
this superbubble at velocities near those expected at tangent points,
including gas that has been swept up sideways from the disk, and
clouds \citep{2007Pidopryhora,2009Pidopryhora}. It is likely some of
the tangent point clouds in the first quadrant region are associated
with this superbubble. 

In \S \ref{GLIMPSE} we suggested that the radial surface density distributions of the disk-halo
clouds closely mirrors the spiral structure of the Galaxy, and in particular the asymmetry 
caused by the Galactic bar.
It is interesting to note that extraplanar gas in some external spiral galaxies, 
such as NGC 4559, appears to be spatially related to star formation activity 
\citep{2005Barbieri}. Also, the presence of extraplanar dust in external galaxies 
implies that if extraplanar gas is a result of feedback, the processes 
transporting the material from the disk must be gentle and likely have low velocities in 
order for the dust grains to survive \citep{2005Howk}. This is consistent with 
the low $\sigma_{cc}$ that we derive for the disk-halo clouds in both quadrants.

Based on the results from our comparisons of the QI and QIV tangent samples, we 
therefore propose 
the following scenario for the origin and evolution of halo \HI\ clouds: the clouds 
are related to areas of star formation, where stellar winds and supernova activity
sweep and push gas from the disk into the lower halo.  Some neutral
gas resides in the walls of superbubbles, whose shells eventually
fragment into clouds. As star formation is abundant in spiral arms,
the clouds are naturally correlated with the spiral structure of the
Galaxy. However, as the timescale for formation of a superbubble
($\sim 20$--$30$~Myr; \citealt{2004deAvillez, 2006McClure-Griffiths})
is large compared to the lifetime of a star-forming region ($\sim
0.1$~Myr; \citealt{2007Prescott}), clouds that are produced in shells
may no longer be at the same locations as the sites where high-mass
stars are forming at the present day.  The scale heights of the cloud population
are reasonable if the gas is brought into the lower halo by
superbubbles or feedback, as high vertical velocities are not required. 
 The magnetic field lines in
a supershell are compressed (e.g., \citealt{2001Ferriere}), which
increases the magnetic pressure and may aid in cloud stability if the
clouds are related to supershells.

It is important to contrast this scenario with that of a standard
galactic fountain, which proposes that clouds are formed by the
cooling and condensing of hot gas that has been expelled from the
disk, which then falls back towards the plane
\citep{1976Shapiro,1980Bregman}. While this model has similarities to
our proposed scenario, an important difference is that the
distribution of clouds would not be expected to have small-scale
features, such as a peaked radial distribution, or a dramatic
difference in the number of clouds between different regions of the
Galaxy at similar radii, as the hot gas from which they condense is
expected to be fairly uniform in the halo \citep{1980Bregman}.  Such
features are clearly present in the disk-halo \HI\ cloud
distributions, which argues for a scenario where clouds are produced
more directly by events occurring within spiral arms.  There is also
no reason to expect the clouds to be associated with loops and
filaments if related to a galactic fountain, but these structures are
often observed. 

In recent years the definition
of a galactic fountain has been expanded to include not just the classical
fountain but any scenario where gas is expelled from the disk into the
lower halo and later returns to the disk regardless of gas phase and
temperature (e.g., see \citealt{2008Spitoni}). Our proposal falls under this broader categorization of a galactic fountain.

\section{Summary Comments}
\label{sec:summary}

A total of 255 disk-halo \HI\ clouds, some $>2$ kpc from the plane,
were detected at the tangent points in the first quadrant region of
GASS data, a region in longitude, latitude and velocity that is
symmetric to the fourth quadrant region studied in Paper~I. Individual
cloud properties in the QI sample are very similar to those in the QIV
sample, having median values $T_{\mathrm{pk}}=0.5$~K, $\Delta
v=10.6$~km~s$^{-1}$, $r=28$~pc and $M_{HI}=700 M_{\odot}$. The clouds
do not have enough mass to be self-gravitating. They must either be
pressure-confined or transitory. The observed increase in linewidth
with distance from the plane suggests that the clouds are
pressure-confined and that the linewidths reflect pressure variations
throughout the halo.

The cloud-cloud line of sight velocity
dispersion is also similar in both regions, with a value $\sigma_{cc}
\approx 16$ km~s$^{-1}$. However, the QI clouds have twice the
exponential scale height as the QIV clouds ($h=800$~pc vs. $h=400$~pc).  
As with the QIV sample, this is many times larger than can be
supported by vertical motions with the magnitude of the cloud-cloud line of sight velocity dispersion.  
Thus the
scale height in the two quadrants is neither derived from nor even
related to the measured velocity dispersion of the cloud population.

Both the cloud numbers and their Galactic distribution are also
markedly different between the two regions, with three times as many
clouds being detected in QI than in QIV.  As the clouds were selected
from a uniform data set using identical criteria, this difference
between the regions must result from a fundamental asymmetry between the
two parts of the Galaxy.  We believe that the differences arise from
the coincidental location of the QI sample on a region where a major
spiral arm merges with the tip of the Galactic bar, whereas the QIV sample
encompasses only a portion of a minor spiral arm.  While there is no
agreement in detail between the Galactic distributions of disk-halo
\HI\ clouds and \HII\ regions, methanol masers, or molecular clouds,
we believe that a link with large-scale star forming regions is the
only explanation for the extreme difference in numbers, distribution
and scale height of clouds in the two regions. 

The most likely scenario is that the disk-halo \HI\
clouds are related to areas of star formation and result from
stellar feedback and superbubbles that have swept gas into the halo forming (or releasing)
the clouds in situ.  These events occur frequently within spiral arms,
but take tens of Myr to reach their maximum extent, by which time the
stellar clusters that produced them are no longer active in star
formation. Simulations of superbubble expansion (e.g.,
\citealt{2008Melioli}; Ford et al., in preparation) and semi-analytic models (e.g.,
\citealt{2008Spitoni}) show that this is a viable mechanism for
producing \HI\ clouds in the lower halo. 

Disk-halo \HI\ clouds are abundant in both QI and QIV, 
a volume of many kpc$^3$. They are not an isolated
phenomenon, but a major component of the Galaxy.  The properties of
the disk-halo cloud population rule out the possibility that most of
the clouds are created through tidal stripping of satellite galaxies or infalling 
primordial gas --- the  clouds are clearly a disk population, concentrated toward the plane, 
highly coupled to Galactic rotation, and correlated with the spiral
structure of the Galaxy.
The disk-halo \HI\ clouds therefore play an important role in Galaxy
evolution and the circulation of gas between the disk and halo, and
are likely common in many external galaxies, though the
angular resolution and sensitivity limits of current instruments would
make their detection difficult.

\acknowledgments

We thank an anonymous referee for comments that led to improvements in the 
presentation and clarity of this work, R.~A. Benjamin for a discussion of the GLIMPSE 
model of the Galaxy, B. Saxton for his help creating Figure \ref{fig:GLIMPSE_image}, 
and J. Bailin for helpful comments on the manuscript.  H.~A.~F. thanks the National 
Radio Astronomy Observatory for support under its Graduate Student Internship Program. 
The National Radio Astronomy Observatory is operated by Associated
Universities, Inc., under a cooperative agreement with the National
Science Foundation.

\bibliography{Paper2_revised.bib}

\end{document}